\documentclass{article}
\usepackage{amssymb,amsmath,graphicx}
\usepackage{amsfonts}
\usepackage[caption=false]{subfig}
\usepackage{epsfig,latexsym,graphicx}
\newtheorem{theorem}{Theorem}[section]

\newtheorem{lemma}[theorem]{Lemma}
\newtheorem{proposition}[theorem]{Proposition}

\newtheorem{remark}{Remark}[section]

\begin{document}

\title{Non-Concave Penalization in Linear Mixed-Effects Models and Regularized Selection of Fixed Effects
}

\author{Abhik Ghosh and Magne Thoresen\\
Department of Biostatistics, University of Oslo\\
abhik.ghosh@medisin.uio.no, magne.thoresen@medisin.uio.no}
\maketitle

\begin{abstract}
Mixed-effect models are very popular for analyzing data with a hierarchical structure, 
e.g. repeated observations within subjects in a longitudinal design, patients nested within centers in a multicenter design. 
However, recently, due to the medical advances, the number of fixed effect covariates collected from each patient can be quite large, 
e.g. data on gene expressions of each patient, and all of these variables are not necessarily important for the outcome.
So, it is very important to choose the relevant covariates correctly for obtaining the optimal inference for the overall study. 
On the other hand, the relevant random effects will often be low-dimensional and pre-specified. 
In this paper, we consider regularized selection of important fixed effect variables in linear mixed-effects models 
along with maximum penalized likelihood estimation of both fixed and random effect parameters based on general non-concave penalties. 
Asymptotic and variable selection consistency with oracle properties are proved for low-dimensional cases as well as for high-dimensionality of non-polynomial order of sample size 
(number of parameters is much larger than sample size). We also provide a suitable computationally efficient algorithm for implementation.  
Additionally, all the theoretical results are proved for a general non-convex optimization problem that applies to several important situations well beyond the mixed model set-up 
(like finite mixture of regressions etc.) illustrating the huge range of applicability of our proposal.
\end{abstract}

\section{Introduction}\label{SEC:intro}

The linear mixed model is a very popular tool for analysis of clustered data from a wide range of applications. 
Relevant examples include, but are not restricted to longitudinal studies and multicenter studies.

Mathematically, let there be $I$ groups of observations, indexed by $i=1, \ldots, I$ and 
assume there are $n_i$ observations in the $i$-th group with total number of observations $n=\sum_{i=1}^I n_i$. 
For each group, we observe the response vector $\boldsymbol{y}_i$ ($n_i$-dimensional), 
the fixed-effect covariates $\boldsymbol{X}_i$ ($n_i\times p$ dimensional) 
and the random-effect covariates $\boldsymbol{Z}_i$ ($n_i\times q$ dimensional, generally a subset of $\boldsymbol{X}_i$).
The model is then given by (Pinheiro and Bates, 2000)
\begin{equation}
	\boldsymbol{y}_i=\boldsymbol{X}_i\boldsymbol{\beta} + \boldsymbol{Z}_i\boldsymbol{b}_i + \boldsymbol{\epsilon}_i,~~~~~i=1,...,I.
	\label{EQ:model1}
\end{equation}
Here, $\boldsymbol{\beta}$ is $p$-dimensional vector of fixed effect coefficients and 
the $\boldsymbol{b}_i$s are the random effects that are assumed to follow a multivariate normal distribution $N_q(0, \boldsymbol{\Psi}_{\boldsymbol{\theta}})$ 
where $\boldsymbol{\theta}$ is the $q^*$ dimensional variance parameter that completely specifies the matrix $\boldsymbol{\Psi}_{\boldsymbol{\theta}}$. 
Also, we assume that the error $\boldsymbol{\epsilon}_i\sim N_{n_i}(0, \sigma^2\boldsymbol{I}_{n_i})$, independent of the random effects $\boldsymbol{b}_i$
and the covariates $\boldsymbol{X}_i$s are independent of $\boldsymbol{\epsilon}_i$s and $\boldsymbol{b}_i$s. 
Note that, for each $i$, given $\boldsymbol{X}_i$ (and $\boldsymbol{Z}_i$),  
$\boldsymbol{y}_i\sim N_{n_i}(\boldsymbol{X}_i\boldsymbol{\beta}, \boldsymbol{V}_i(\boldsymbol{\theta},\sigma^2))$,
where $\boldsymbol{V}_i(\boldsymbol{\theta},\sigma^2)=\boldsymbol{Z}_i\boldsymbol{\Psi}_{\boldsymbol{\theta}}\boldsymbol{Z}_i^T + \sigma^2\boldsymbol{I}_{n_i}$.

In the example of a multicenter study, the centers are referred to as the groups and we have $n_i$ patients treated in the $i$-th center. 
The random effect covariates $\boldsymbol{Z}_i$s are then factors related to the centers and are generally few. 
However, modern medical studies gather lots of information about each patient, e.g. high-dimensional genomic measurements. 
Hence, the pool of fixed effects covariates $\boldsymbol{X}_i$ can be quite large. 
All of these variables are not necessarily important to study the effect of treatment or any other response variable we are studying, 
so variable selection becomes an issue.

For this reason, variable selection in the mixed effect models has become a very important research topic in recent literature. 
Although there are several classical works on the estimation and testing under linear and generalized linear mixed effect models, 
efficient variable selection procedures under this set-up has gained attention more recently. 
Vaida and Blanchard (2005) and Liang, Wu and Zou (2008) proposed and studied the conditional AIC approach for variable selection in mixed effect models
and described the concepts of degrees of freedom in detail. Chen and Dunson (2003) have considered Bayesian variable selection for the 
random effects in linear mixed-effect models and Pu and Niu (2006) have extended the general information criterion 
to choose the fixed effects under similar set-up.
Bondell, Krishna and Ghosh (2010), Ibrahim et al.~(2011) and Lin , Pang and Jiang (2013) considered the simultaneous selection of fixed and random effects 
through different approaches which are applicable mainly to situations where there are many random effect variables 
along with the large pool of fixed effect variables. 
However, as mentioned above, in most applications in medical and clinical biology, 
the number of random effects is generally small and can be considered pre-fixed, 
and we are mainly interested in selecting the fixed effects from a large pool of possible candidates.  
There are only a few approaches to variable selection under such situations, e.g. Taylor et al.~(2012), Xu et al.~(2015) etc.;
most of these approaches relate to the least absolute shrinkage and selection operator (LASSO, Tibshirani, 1979)
or its extension involving some generalization of the $L_1$ penalty.
However, all the works mentioned till now are limited to the classical low-dimensional set-up 
with the number of available observations ($n$) being more than the total number of parameters ($P=p+q^*+1$) in the model
and hence they fail in case of modern high-dimensional data-sets with $P\gg n$.
More recently, only the Lasso approach with $L_1$ penalty has been extended to such high-dimensional set-up by Schelldorfer, Buhlmann and Van de Geer (2011)
and its numerical, computational aspects and applications have been discussed in Fazli et al.~(2011), 
Rohart, San Cristobal and Laurent (2014), Jakubik (2015) and Bonnet, Gassiat and Levy-Leduc (2015).
Muller, Scealy and Welsh (2013) have provided a good review of these variable selection methods.

However, recent advances in variable selection under the regression set-up show several advantages of 
using more general non-concave penalty functions over the classical $L_1$ penalty based methods.
In a pioneer paper by Fan and Li (2001), a smoothly clipped absolute deviation (SCAD) penalty has been proposed in case of regression models
while discussing the non-concave penalized likelihood under classical low-dimensional set-up ($P<n$). 
The general theory of non-concave penalty based variable selection has also been extended to the cases of diverging number of parameters and 
to the high-dimensional regression set-up by Fan and Peng (2004) and Fan and Lv (2011) respectively.
All these papers illustrate useful variable selection properties of the general non-concave penalty, specially the SCAD penalty, 
over the $L_1$ penalty used in the Lasso based approaches under the regression set-up. 
In particular, it has been shown that the SCAD penalty reduces the number of false significant variables chosen compared to the Lasso approaches
and satisfies both the $\sqrt{n}$-consistency and oracle properties of variable selection which cannot be achieved simultaneously by the $L_1$-penalty in Lasso. 
Such improvements can also be expected to be achieved for variable selection in linear mixed effect models 
by considering a general non-concave penalized likelihood approach. 
However, such non-concave penalty under the mixed-model set-up has been considered only by Fan and Li (2012), 
where a sequential selection of the fixed and random effect variables is considered.
One major drawback of their approach is that they just used some proxy matrix with penalized profile likelihood to avoid the
unknown covariance matrix of random effects without estimating them. Although they have provided some criterion to choose the proxy matrix, 
it is quite difficult to understand which proxy to choose in any given practical situation;
furthermore, the simultaneous estimation of fixed and random effect parameters is also important in addition to selecting the relevant variables. 

In this paper, we consider the penalized likelihood based estimation of the fixed and random effect parameters simultaneously 
using general non-concave penalties along with a regularized selection of important fixed effect variables. 
Thus, our proposal will avoid the practical complication of the proxy matrix of the Fan and Li (2012) approach 
and reduce computational efforts by generating the random effect estimates also in the same stage avoiding the two step process. 
Indeed, we consider the penalized likelihood for both the fixed and the random effect parameters ($\boldsymbol{\beta}$, $\boldsymbol{\theta}$, $\sigma^2$)  
with general non-convex penalties and maximize it simultaneously to get their maximum penalized likelihood estimators (MPLEs). 
The regularized selection of the fixed effect variables has been considered via a suitable, computationally efficient algorithm
and their consistency and oracle properties are proved for the classical low-dimensional cases ($P<n$) as well as 
for high-dimensionality of non-polynomial order of sample size ($P\gg n$ with $\log P =O(n^\alpha)$ for some $\alpha \in (0,1))$.
%
The main contribution of the paper can be summarized as follows:

Instead of deriving the properties of the MPLEs only for the linear mixed-model (\ref{EQ:model1}),
the paper provides a general asymptotic theory with nice optimality results	for penalized maximum likelihood estimation
based on general non-convex loss functions and general non-concave penalties. 
The simplification for the linear mixed-effect model (\ref{EQ:model1}) has also been provided as an illustrations. 
This general set-up includes several non-standard	statistical models like finite mixture of regressions etc., 
besides our mixed effect models, and hence extends the scope of the paper. 
We believe such general asymptotic results contribute importantly to the literature, 
since all the previously existing results were only for convex loss or 
for some limited specific models having non-convex loss with a specific penalty.

Further, the general asymptotic theory, in particular the asymptotic consistency and variable selection oracle property, 
has been developed for the classical low-dimensional cases with $P<n$ 
as well as for the modern high-dimensional set-up where the number of parameters increases exponentially with the sample size. 
Under the linear mixed model (\ref{EQ:model1}), the asymptotic distribution of the penalized estimators 
with a general class of non-concave penalty functions has also been provided under high-dimensionality,
which is another interesting addition to the literature, 
as there are no existing result on the asymptotic distribution of the MPLEs under the high-dimensional mixed model even with $L_1$ penalty. 

From an application point-of-view, the paper also illustrates that, in a linear mixed model,  
the use of the SCAD penalty yields better results in terms of MSE and false positives for 
the estimation and selection of the fixed effect variables respectively, compared to the classical $L_1$ penalty. 
Although there are some existing works with some specific penalty for the linear mixed model with low-dimensional set-up, 
the advantages of SCAD is a major contribution of the current paper. 
On the other hand, there are only a few approaches of penalized estimation in the high-dimensional linear mixed-effect model
and our proposal with SCAD clearly outperforms them for estimation and selection of the fixed effect parameters. 
This motivates and provides guidelines for practitioners to use the appropriate penalty for any real-life application following the present work.

The rest of the paper is organized as follows: In Section \ref{SEC:penalized_estimation} we will describe the
procedure of the general penalized likelihood estimation with non-concave penalty functions 
along with intuitions behind their constructions and penalty used. 
In Section \ref{SEC:Theory} we will present the main theoretical results and Section \ref{SEC:computation} 
will consider the computational aspect of the proposal, illustrating suitable numerical solutions for the problem.
Appropriate simulations and real data illustrations have been provided in Section \ref{SEC:numerical}
and the paper ends with some concluding remarks in Section \ref{SEC:Conclusion}.
For simplicity in presentations, proofs of all the results have been moved to the Appendix.

\section{Penalized Likelihood based Estimation and Variable Selection}\label{SEC:penalized_estimation}

\subsection{General Non-Concave Penalty Functions}\label{SEC:penalty}

The penalty function is an important component of regularized variable selection, 
which largely determines the properties of the resulting penalized estimators and variable selection. 
Let us denote the penalty function at a scalar parameter $\gamma$ as $p_\lambda(|\gamma|)$,
where $\lambda$ is a tuning parameter that controls the amount of regularization.
Many penalty functions have been used for regularization in existing literature. These includes the
popular $L_1$ penalty $p_\lambda(|\gamma|)=\lambda|\gamma|$, the $L_2$ penalty $p_\lambda(|\gamma|)=\lambda|\gamma|^2$
or, more generally the bridging $L_q$ penalty $p_\lambda(|\gamma|)=\lambda|\gamma|^q$ for $q\in (0,2)$ 
(Frank and Friedman, 1993; Fu, 1998, Knight and Fu, 2000).
In the simplified penalized least square problem, where we minimize $\frac{1}{2}(z-\gamma)^2 + p_\lambda(|\gamma|)$ based on given data $z$,
the resulting solution for $\gamma$ is the LASSO (Tibshirani, 1996) for $L_1$ penalty and ridge regression for the $L_2$ penalty.
Using the $L_0$ penalty leads to the method of best subset selection whereas the hard thresholding penalty 
\begin{equation}
p_\lambda(|\gamma|)= \lambda^2 - (|\gamma|-\lambda)^2 I(|\gamma|<\lambda)
\label{EQ:Hard}
\end{equation}
of Antoniadis (1997) and Fan (1997) yields the solution  $\hat\gamma(z) = zI(|z|>\lambda)$. 
Another popular penalty, proposed by Fan (1997) in the context of wavelength analysis,
is the smoothly clipped absolute deviation (SCAD) penalty,  defined through its derivative
\begin{equation}
p_\lambda'(|\gamma|) = \lambda\left\{I(|\gamma|\leq\lambda) + \frac{(a\lambda-|\gamma|)_+}{(a-1)\lambda}I(|\gamma|>\lambda)\right\},
\label{EQ:SCAD}
\end{equation}
for some $a>2$ which leads to the solution 
\begin{equation}
\hat\gamma(z) = \lambda\left\{ 
\begin{array}{lcl}
sgn(z)(|z|-\lambda)_+  & \mbox{ if } & |z| \leq 2\lambda,\\
\frac{1}{(a-2)}\left[(a-1)z - sgn(z) a \lambda\right]  & \mbox{ if } & 2\lambda \leq |z| \leq a\lambda,\\
z  & \mbox{ if } & |z| > a\lambda.
\end{array}\right.
\label{EQ:SCAD_sol}
\end{equation}

Fan and Li (2001) characterized a good penalty function based on three properties: (i) Unbiasedness to avoid unnecessary modeling biases,
(ii) Sparsity in order to get automatic regularized selection of important variables, and 
(iii) Continuity of the resulting estimator in data to avoid prediction instability of the model. 
Following Fan and Li (2001) and Antoniadis and Fan (2001), 
sufficient conditions for a penalty $p_\lambda(|\gamma|)$ to satisfy the above three properties respectively are 
(i) $p_\lambda'(|\gamma|)=0$ for large $\gamma$, (ii) the minimum of $(|\gamma|+p_\lambda'(|\gamma|))$ is positive,
and (iii) the minimum of $(|\gamma|+p_\lambda'(|\gamma|))$ is attained at 0. 
In particular, the $L_q$ penalty with $q>1$ (including the $L_2$ penalty) provides shrinkage in the solution but do not satisfy the sparsity property.
On the other hand, $L_q$ penalty with $q\leq 1$ (including the $L_1$ penalty) satisfies the sparsity property 
but fails to satisfy the unbiasedness requirement due to excessive penalization at the large parameter values.
Further, the hard thresholding penalty  results in a solution that is not continuous in data.
However, the SCAD penalty satisfies all the three desired properties and seems to be the most useful candidate 
for regularized variable selection over the most popular choice of $L_1$ penalty.

\subsection{The Maximum Penalized Likelihood Estimation}\label{SEC:MPLE}

Let us consider the linear mixed effect model given in (\ref{EQ:model1}). 
We will first define the maximum likelihood estimator of the parameters $(\boldsymbol{\beta}, \boldsymbol{\eta})$ under penalization
where $\boldsymbol{\eta} =(\sigma, \boldsymbol{\theta})$ is the variance parameters in the model.
Since $\boldsymbol{y}_i\sim N_{n_i}(\boldsymbol{X}_i\boldsymbol{\beta}, \boldsymbol{V}_i(\boldsymbol{\theta},\sigma^2))$ for each $i$, 
the log-likelihood function is given by 
\begin{eqnarray}
l_n(\boldsymbol{\beta},\boldsymbol{\eta}) &=& -\frac{1}{2}\sum_{i=1}^I \left[n_i\log(2\pi) + \log|\boldsymbol{V}_i(\boldsymbol{\theta},\sigma^2)| 
+ (\boldsymbol{y}_i-\boldsymbol{X}_i\boldsymbol{\beta})^T\boldsymbol{V}_i(\boldsymbol{\theta},\sigma^2)^{-1}
(\boldsymbol{y}_i - \boldsymbol{X}_i\boldsymbol{\beta})\right]
\nonumber\\
&=& -\frac{1}{2}\left[n\log(2\pi) + \log|\boldsymbol{V}(\boldsymbol{\theta},\sigma^2)| 
+ (\boldsymbol{y}-\boldsymbol{X}\boldsymbol{\beta})^T\boldsymbol{V}(\boldsymbol{\theta},\sigma^2)^{-1}(\boldsymbol{y}-\boldsymbol{X}\boldsymbol{\beta})\right],
\label{EQ:log-likelihood}
\end{eqnarray}
where $\boldsymbol{y}=(\boldsymbol{y}_1^T, \ldots, \boldsymbol{y}_I^T)^T$, 
$\boldsymbol{X}=(\boldsymbol{X}_1^T, \ldots, \boldsymbol{X}_I^T)^T$ and 
$\boldsymbol{V}={\rm Diag}\{\boldsymbol{V}_1, \ldots, \boldsymbol{V}_I\}$ are the stacked matrices. 

Note that we have assumed that the random effects are pre-specified and we only want to select important fixed effects 
through regularized penalization. 
Let us consider the general class of non-negative penalty functions $P_{n,\lambda}(\cdot)$ 
that might depend on the sample size $n$ along with the regularization parameter $\lambda$.
Generally, in practice, this dependence comes through the dependence of $\lambda$ on $n$; 
for example $P_{n,\lambda}(\cdot) = n p_{\lambda_n}(\cdot)$ with $p_\lambda$ being any penalty function defined in the previous subsection.
Then, we consider the minimization of the following penalized negative log-likelihood objective function:
\begin{eqnarray}
Q_{n,\lambda}(\boldsymbol{\beta},\boldsymbol{\eta}) &=& -l_n(\boldsymbol{\beta},\boldsymbol{\eta}) + \sum_{j=1}^p p_\lambda(|\beta_j|)
\nonumber\\
&=& \frac{1}{2}\left[n\log(2\pi) + \log|\boldsymbol{V}(\boldsymbol{\theta},\sigma^2)| 
+ (\boldsymbol{y}-\boldsymbol{X}\boldsymbol{\beta})^T\boldsymbol{V}(\boldsymbol{\theta},\sigma^2)^{-1}(\boldsymbol{y}-\boldsymbol{X}\boldsymbol{\beta})\right] 
+ \sum_{j=1}^p P_{n,\lambda}(|\beta_j|).
\label{EQ:penal_log-likelihood}
\end{eqnarray}
The minimization of $Q_{n,\lambda}(\boldsymbol{\beta},\boldsymbol{\eta}) $ provides the MPLE of 
$(\boldsymbol{\beta},\boldsymbol{\eta})$ with regularization parameter $\lambda$ and 
can simultaneously select the important components of $\boldsymbol{\beta}$ for appropriately chosen penalty functions.
Note that this minimization is not a convex optimization problem since the log-likelihood is convex only with respect to $\boldsymbol{\beta}$ 
and non-convex with respect to $\boldsymbol{\eta}$. So, we cannot simply use the techniques of  convex optimization to obtain the MPLEs.
We will discuss some suitable quadratic approximations and iterative algorithms to solve this non-convex optimization problem in Section \ref{SEC:computation}

%
%
%
%


\section{Theoretical Results: Consistency and Oracle Property}\label{SEC:Theory}

We first consider a more general non-convex optimization problem, where we have to minimize the general objective function 
\begin{eqnarray}
Q_{n,\lambda}(\boldsymbol{\beta},\boldsymbol{\eta}) &=& L_n(\boldsymbol{\beta},\boldsymbol{\eta}) + \sum_{j=1}^p P_{n,\lambda}(|\beta_j|)
\label{EQ:penal_GenObj}
\end{eqnarray}
with respect to the parameters $(\boldsymbol{\beta}, \boldsymbol{\eta})$ for a general loss function 
$L(\boldsymbol{\beta},\boldsymbol{\eta})$ defined based on observations $\boldsymbol{V}_k$ for $k=1,\ldots, n$. 
Keeping consistent with our mixed model set-up, we will assume that the general loss function  $L(\boldsymbol{\beta},\boldsymbol{\eta})$
is also convex only in $\boldsymbol{\beta}$ and non-convex in $\boldsymbol{\eta}$. 
Note that, it corresponds to the objective function (\ref{EQ:penal_log-likelihood}) for the choice $L_n(\boldsymbol{\beta},\boldsymbol{\eta})$ being the negative log-likelihood 
of the mixed model given by (\ref{EQ:log-likelihood}) and $\{\boldsymbol{V}_k\}=\{(y_{ij}, X_{ij})\}$ with $\boldsymbol{Z}_i \subset \boldsymbol{X}_i$.
Such general non-convex optimization problems occur in many other important situations besides our linear mixed model; 
for example the finite mixture regression model as considered in Stadler, Buhlmann and van de Geer (2010) also have an objective function of exactly the same type.

We will first develop theoretical results for this general objective function in two situations -- 
(i) fixed number of parameters with small $p<n$ and 
(ii) high-dimensionality of non-polynomial (NP) order with $\log p =O(n^\alpha)$ for some $\alpha \in (0,1)$. 
The properties of the MPLE under the mixed model set-up will then be described as special cases of the general results.

\subsection{General Non-convex likelihood with fixed number of parameters}\label{SEC:Theory_fixed}

We will assume that the observations $\boldsymbol{V}_k$ are independent and identically distributed with a density 
$f(\boldsymbol{V}; \boldsymbol{\beta}, \boldsymbol{\eta})$
and we want to estimate the parameters $(\boldsymbol{\beta},\boldsymbol{\eta})$ 
by minimizing the general non-convex objective function (\ref{EQ:penal_GenObj}). 
Here $\boldsymbol{\beta}$ is a $p$ dimensional vector which we want to select by regularization and 
$\boldsymbol{\eta}$ is a $d$-dimensional vector of parameters that are outside the scope of regularized selection 
(these corresponds to the pre-fixed variance parameters in the linear mixed model with $d=q^*+1$).
Suppose $(\boldsymbol{\beta}_0,\boldsymbol{\eta}_0)$ is the true parameter value generating the observed data $\{\boldsymbol{V}_k\}$.
Consider the general negative likelihood loss
$L_n(\boldsymbol{\beta},\boldsymbol{\eta})= - \sum_{i=1}^n \log f(\boldsymbol{V}_i; \boldsymbol{\beta},\boldsymbol{\eta})$ 
which is assumed to be convex in $\boldsymbol{\beta}$ but non-convex in $\boldsymbol{\eta}$.
Let us assume some regularity conditions on the model; these are standard requirements of asymptotic derivations 
for general maximum likelihood estimators (Lehmann, 1983) and are satisfied by most common statistical models.

\bigskip
\noindent
\textbf{Assumptions on the model:}
\begin{itemize}
\item[(MA1)] The model is identifiable and the support of $f(\boldsymbol{V}; \boldsymbol{\beta}, \boldsymbol{\eta})$ 
is independent of the parameter $(\boldsymbol{\beta}, \boldsymbol{\eta})$.
Further, the density $f(\boldsymbol{V}; \boldsymbol{\beta}, \boldsymbol{\eta})$ possesses first and second order derivatives satisfying 
$$
E\left[\frac{\partial \log f(\boldsymbol{V}; \boldsymbol{\beta}, \boldsymbol{\eta})}{\partial(\boldsymbol{\beta}, \boldsymbol{\eta})}\right] = 0,
$$ 
and 
$$
\boldsymbol{I}(\boldsymbol{\beta},\boldsymbol{\eta}) 
= E\left[\left(\frac{\partial \log f(\boldsymbol{V}; \boldsymbol{\beta}, \boldsymbol{\eta})}{\partial(\boldsymbol{\beta}, \boldsymbol{\eta})}\right)
\left(\frac{\partial \log f(\boldsymbol{V}; \boldsymbol{\beta}, \boldsymbol{\eta})}{\partial(\boldsymbol{\beta}, \boldsymbol{\eta})}\right)^T\right] 
= E\left[- \frac{\partial^2 \log f(\boldsymbol{V}; \boldsymbol{\beta}, \boldsymbol{\eta})}{\partial(\boldsymbol{\beta}, \boldsymbol{\eta})^2}\right]. 
$$ 
\item[(MA2)] The Fisher information matrix $\boldsymbol{I}(\boldsymbol{\beta}, \boldsymbol{\eta})$ is finite and positive definite 
at $(\boldsymbol{\beta}_0,\boldsymbol{\eta}_0)$ .

\item[(MA3)] There exists an open subset of the parameter space containing the true parameters on which the density 
$f(\boldsymbol{v};\boldsymbol{\beta},\boldsymbol{\eta})$ admits all its third order partial derivatives 
for almost all $\boldsymbol{v}$ which are uniformly (on that open subset) bounded 
by some functions having finite expectation under the true parameter value.
\end{itemize}

For this case of fixed number of parameters, we will choose the penalty as $P_{n,\lambda}(\cdot) = n p_{\lambda_n}(\cdot)$
and define
\begin{equation}
a_n = \max \{ p_{\lambda_n}'(|\beta_{0j}|) : \beta_{0j} \neq 0\} ,~~b_n = \max \{ p_{\lambda_n}''(|\beta_{0j}|) : \beta_{0j} \neq 0\}
\end{equation}
Then, we also need the following assumptions on the penalty function.

\bigskip
\noindent
\textbf{Assumptions on the penalty:}
\begin{itemize}
\item[(PA1)]$ b_n\rightarrow 0$ as $n\rightarrow\infty$.

\item[(PA2)] $\liminf_{n\rightarrow\infty}\liminf_{\beta\downarrow 0+}\frac{p_{\lambda_n}'(\theta)}{\lambda_n} >0.$
\end{itemize}

These conditions hold for the usual penalty functions under suitable assumptions on the regularization sequence $\lambda_n$.
Further, we will assume that the true parameter value $\boldsymbol{\beta}_0$ of $\boldsymbol{\beta}$ is sparse and given by
$$
\boldsymbol{\beta}_0=(\beta_{01},\ldots, \beta_{0p})^T = (\boldsymbol{\beta}_0^{(1)T}, ~\boldsymbol{\beta}_0^{(2)T})^T
$$
where $\boldsymbol{\beta}_0^{(1)}$ is of dimension $s << p$ and $\boldsymbol{\beta}_0^{(2)}= 0_{p-s}$, the $(p-s)$-dimensional vector of all entries zero. 
Let $\boldsymbol{\beta}=(\boldsymbol{\beta}^{(1)T}, ~\boldsymbol{\beta}^{(2)T})$ 
denote the corresponding partitioning of the general parameter vector $\boldsymbol{\beta}$, where $\boldsymbol{\beta}^{(2)}$ is not necessarily zero. 
Our first theorem shows the existence of a penalized estimator (PE) of $(\boldsymbol{\beta},\boldsymbol{\eta})$ 
as the minimizer of the general objective function (\ref{EQ:penal_GenObj}) 
that converges to the true value at the rate $O_p(n^{-1/2} + a_n)$ for $\boldsymbol{\beta}$ and at the rate $O_p(n^{-1/2})$ for $\boldsymbol{\eta}$.
This shows the $\sqrt{n}$-consistency of the PE whenever the penalty is chosen to satisfy $a_n\rightarrow 0$, 
which holds for hard thresholding and SCAD penalty with $\lambda_n\rightarrow 0$.
For the $L_1$ penalty, however, we have $a_n=\lambda_n$ and hence we need to have $\lambda_n=O(n^{-1/2})$
to achieve $\sqrt{n}$-consistency of the $L_1$ penalized estimator of $\boldsymbol{\beta}$, as seen in the case of the Lasso (Fan and Li, 2001).

\begin{theorem}
Consider the above mentioned general set-up with Assumptions (MA1)--(MA3). If the penalty function satisfies Assumption (PA1),
then there exists a local minimizer $(\hat{\boldsymbol{\beta}}, \hat{\boldsymbol{\eta}})$ of $Q_{n,\lambda}(\beta,\eta)$ satisfying 
\begin{equation}
||\hat{\boldsymbol{\beta}} - \boldsymbol{\beta}_0|| = O_p(n^{-1/2} + a_n),~~~||\hat{\boldsymbol{\eta}} - \boldsymbol\eta_0|| = O_p(n^{-1/2}).
\label{EQ:results_3.1}
\end{equation}
\label{THM:fixed_consistency}
\end{theorem}

Our next theorem presents the oracle property by showing the sparsity of the local minimizer in Theorem \ref{THM:fixed_consistency}
and also presents the asymptotic distribution of the non-zero elements of $\hat{\boldsymbol{\beta}}$ and of $\hat{\boldsymbol{\eta}}$.
Unlike many other existing proposals, this asymptotic distribution helps us to estimate the standard error of the sparse estimate of $\boldsymbol\beta$ 
as well as the variance parameters $\boldsymbol\eta$.

\begin{theorem}
Consider the above mentioned general set-up with Assumptions (MA1)--(MA3) and (PA2). 
If $\lambda_n\rightarrow 0$ and $\sqrt{n}\lambda_n \rightarrow \infty$ as $n\rightarrow\infty$, 
then the local minimizer   $(\hat{\boldsymbol{\beta}}, \hat{\boldsymbol{\eta}}) 
= ((\hat{\boldsymbol{\beta}}^{(1)T},\hat{\boldsymbol{\beta}}^{(2)T})^T, \hat{\boldsymbol{\eta}})$ in Theorem \ref{THM:fixed_consistency}
satisfies $\hat{\boldsymbol{\beta}}^{(2)} = 0$ with probability tending to one and
\begin{eqnarray}
&& \sqrt{n}(\boldsymbol{I}_{1}(\boldsymbol\beta_{0}^{(1)}, \boldsymbol\eta_0) + \boldsymbol\Sigma)
\left\{  \hat{\boldsymbol{\beta}}^{(1)} - \boldsymbol\beta_0^{(1)} + (\boldsymbol{I}_{1}(\boldsymbol\beta_{0}^{(1)}, \boldsymbol\eta_0) 
+ \boldsymbol\Sigma)^{-1}\boldsymbol\zeta\right\}
\rightarrow^\mathcal{D} N_s(0, \boldsymbol{I}_{1}(\boldsymbol\beta_{0}^{(1)}, \boldsymbol\eta_0)) \label{EQ:Asymp_beta_fixed}\\
&& \sqrt{n}(\hat{\boldsymbol{\eta}} - \boldsymbol\eta_0) \rightarrow^\mathcal{D} 
N_d(0, \boldsymbol{I}_{2}(\boldsymbol\beta_{0}^{(1)}, \boldsymbol\eta_0)^{-1}), \label{EQ:Asymp_eta_fixed}
\end{eqnarray}
where 
$$
\boldsymbol\Sigma={\rm Diag}\{p_{\lambda_n}''(|\beta_{01}|), \ldots, p_{\lambda_n}''(|\beta_{0s}|) \},~~~~
\boldsymbol\zeta=(p_{\lambda_n}'(|\beta_{01}|)sgn(\beta_{01}), \ldots, p_{\lambda_n}'(|\beta_{0s}|)sgn(\beta_{0s}))^T,
$$
and $\boldsymbol{I}_{1}(\boldsymbol\beta_{0}^{(1)}, \boldsymbol\eta_0)$ and $\boldsymbol{I}_{2}(\boldsymbol\beta_{0}^{(1)}, \boldsymbol\eta_0)$ 
are the Fisher information matrices corresponding to $\beta^{(1)}$ and $\eta$ respectively
assuming $\boldsymbol\beta^{(2)}=0$.
\label{THM:fixed_oracle}
\end{theorem}

From the above theorem, we can easily obtain the asymptotic covariance matrices of $(\hat{\boldsymbol{\beta}}_{0}^{(1)}, \hat{\boldsymbol{\eta}}_0)$ 
and provide a sandwich estimator of the asymptotic variance of the estimators of $\boldsymbol\beta$ and $\boldsymbol\eta$
as given by
\begin{eqnarray}
\widehat{Cov}(\hat{\boldsymbol{\beta}}_1) &=& \frac{1}{n}\left[\nabla_\beta^2 L(\hat{\boldsymbol{\beta}}_1,\hat{\boldsymbol{\eta}}_1) + \Sigma(\hat{\boldsymbol{\beta}}_1)\right]^{-1}
\widehat{Cov}\left\{\nabla_\beta^2 L(\hat{\boldsymbol{\beta}}_1,\hat{\boldsymbol{\eta}}_1)\right\}
 \left[\nabla_\beta^2 L(\hat{\boldsymbol{\beta}}_1,\hat{\boldsymbol{\eta}}_1) + \Sigma(\hat{\boldsymbol{\beta}}_1)\right]^{-1},
\label{EQ:est_SD_beta}\\
\widehat{Cov}(\hat{\boldsymbol{\eta}}_1) &=& \frac{1}{n} \left[\nabla_\eta^2 L(\hat{\boldsymbol{\beta}}_1,\hat{\boldsymbol{\eta}}_1)\right]^{-1}.
\label{EQ:est_SD_eta}
\end{eqnarray}

%

\subsection{General Non-convex Loss with high (NP) dimensionality }\label{SEC:Theory_high}

Consider the high-dimensional set-up where $p$ is of non-polynomial (NP) order of sample size ($n$), i.e., 
$\log p=O(n^\alpha)$ for some $\alpha \in (0,1)$. In this section we will consider the general non-smooth loss function 
$L_n(\boldsymbol\beta,\boldsymbol\eta)$, which is convex in $\boldsymbol\beta$ but non-convex in $\boldsymbol\eta$. 
We will prove the oracle consistency and variable selection optimality of our proposed set-up 
under this high-dimensional set-up. Consider the following assumptions:

\bigskip
\noindent
\textbf{Assumptions on the penalty (P):}\\
The general penalty function $P_{n,\lambda}(t): [0, \infty) \rightarrow \mathbb{R}$ satisfies
\begin{itemize}
\item[(i)] $P_{n,\lambda}(0) = 0$

\item[(ii)] $P_{n,\lambda}(t)$ is concave and non-decreasing on $[0,\infty)$ and has continuous derivative $P_{n,\lambda}'(t)$ on $(0,\infty)$

\item[(iii)] $\sqrt{s} P_{n,\lambda}'(d_n) = o(d_n)$, where $s$ is the number of non-zero elements of $\boldsymbol\beta$ and 
$$d_n=\frac{1}{2}\min\{|\beta_{0j}|:\beta_{0j}\neq 0, ~j=1, \ldots, p\}
$$  denotes the strength of the signal

\item[(iv)] There exists a constant $c>0$ such that $\sup_{\boldsymbol\beta\in B(\boldsymbol\beta_{S_0}, cd_n)} \zeta(\boldsymbol\beta) = o(1)$, where
\begin{equation}
\zeta(\boldsymbol\beta) = \limsup_{\epsilon \rightarrow 0+} \max_{j\leq s} \sup_{t_1<t_2: (t_1,t_2)\in (|\beta_j|-\epsilon, |\beta_j|+\epsilon)}
- \left[\frac{P_{n,\lambda}(t_2) - P_{n,\lambda}(t_1)}{t_2-t_1}\right].
\label{EQ:zeta_beta}
\end{equation}
\end{itemize}

These assumptions are exactly the same as Assumption 4.1 in Fan and Liao (2014),
used first for penalized estimation in endogenous regression model with some general non-smooth loss function.
It is easy to verify these assumptions for the standard $L_q$ penalty with $q\leq 1$, hard thresholding and the SCAD penalty 
for a properly chosen regularization parameter $\lambda$.

Now, let us define the oracle space $\mathcal{B} = \{\boldsymbol\beta\in \mathbb{R}^p ~:~ \beta_j =0 \mbox{ for } j \notin S\}$.
For $\boldsymbol\beta=(\boldsymbol\beta_S^T, 0)^T\in \mathcal{B}$, let us denote 
$L_1(\boldsymbol\beta_S, \boldsymbol\eta) = L_n((\boldsymbol\beta_S^T, 0)^T,\boldsymbol\eta)$. 
Also let $\nabla_S L_1(\boldsymbol\beta_S, \boldsymbol\eta) = \frac{\partial}{\partial\boldsymbol\beta_S}L_n((\boldsymbol\beta_S^T, 0)^T,\boldsymbol\eta)$,
$\nabla_S^2 L_1(\boldsymbol\beta_S, \boldsymbol\eta) 
= \frac{\partial^2}{\partial\boldsymbol\beta_S\partial\boldsymbol\beta_S^T}L_n((\boldsymbol\beta_S^T, 0)^T,\boldsymbol\eta)$, 
$\nabla_\eta L_1(\boldsymbol\beta_S, \boldsymbol\eta) = \frac{\partial}{\partial\boldsymbol\eta}L_n((\boldsymbol\beta_S^T, 0)^T,\boldsymbol\eta)$, 
$\nabla_\eta^2 L_1(\boldsymbol\beta_S, \boldsymbol\eta) 
= \frac{\partial^2}{\partial\boldsymbol\eta\partial\boldsymbol\eta^T}L_n((\boldsymbol\beta_S^T, 0)^T,\boldsymbol\eta)$,
$\nabla_{S\eta} L_1(\boldsymbol\beta_S, \boldsymbol\eta) 
= \frac{\partial^2}{\partial\boldsymbol\beta_S\partial\boldsymbol\eta^T}L_n((\boldsymbol\beta_S^T, 0)^T,\boldsymbol\eta)$ 
and 
$\nabla^2 L_1(\boldsymbol\beta_S, \boldsymbol\eta) = \begin{pmatrix}
\nabla_S^2 L_1(\boldsymbol\beta_S, \boldsymbol\eta) & \nabla_{S\eta} L_1(\boldsymbol\beta_S, \boldsymbol\eta)\\
\nabla_{S\eta} L_1(\boldsymbol\beta_S, \boldsymbol\eta)^T & \nabla_\eta^2 L_1(\boldsymbol\beta_S, \boldsymbol\eta)
\end{pmatrix}.
$
Then, we consider the following assumptions on the model based loss function:

\bigskip
\noindent
\textbf{Assumptions on the loss function (L1):}\\
$L_n(\boldsymbol\beta_S, 0; \boldsymbol\eta) $ is twice differentiable with respect to $\boldsymbol\beta_S$ and $\boldsymbol\eta$ 
in the neighborhood of true values $(\beta_{S0}, 0; \boldsymbol\eta_0)$
and there exists sequences of positive reals $a_n=o(d_n)$ and $c_n=o(1)$ such that the following are satisfied:
\begin{itemize}
\item[(i)] $||\nabla_S L_n(\boldsymbol\beta_{S0}, 0; \boldsymbol\eta_0)|| = O_p(a_n)$ and  $||\nabla_\eta L_n(\beta_{S0}, 0; \boldsymbol\eta_0)|| = O_p(c_n)$

\item[(ii)] For any $\epsilon>0$, there exists some positive constant $C_\epsilon$ such that 
$$
P\left(\lambda_{min}(\nabla^2 L_n(\boldsymbol\beta_{S0}, 0; \boldsymbol\eta_0))> C_\epsilon\right) > 1-\epsilon, ~~~~~\mbox{ for all large } n
$$
 
\item[(iii)] For any given $\epsilon >0$, $\delta>0$ and non-negative sequences $\alpha_n=o(d_n)$ and $\gamma_n=o(1)$, 
there exist a large $N^*$ such that 
$$
P\left(\sup_{||\boldsymbol\beta_S - \boldsymbol\beta_{S0}||\leq\alpha_n, ||\boldsymbol\eta - \boldsymbol\eta_{0}||\leq\gamma_n}
||\nabla^2 L_n(\boldsymbol\beta_{S}, 0; \boldsymbol\eta) - \nabla^2 L_n(\boldsymbol\beta_{S0}, 0; \boldsymbol\eta_0)|| \leq \delta \right) 
> 1-\epsilon, ~~~\mbox{ for all } n>N^*.
$$
\end{itemize}

These assumptions are straightforward extension of the corresponding assumptions in the low-dimensional case and 
can be shown to be satisfied by the likelihood loss for common statistical models. 
We will illustrate them for the linear mixed effect model under consideration in Section \ref{SEC:Theory_LMM}.
However, we would like to emphasis that these assumptions indeed apply to completely general loss functions (which need not to be even smooth) 
and hence the results  obtained below can be applied to several more general problems as well.

\begin{theorem}[Oracle consistency]
Under Assumptions (P) and (L1), there exists a local minimum $(\hat{\boldsymbol{\beta}}=(\hat{\boldsymbol{\beta}}_S^T, 0)^T, ~\hat{\boldsymbol{\eta}})$ of
$$
Q_{n,\lambda}(\boldsymbol\beta_S, 0; \boldsymbol\eta)  = L_n(\boldsymbol\beta_S, 0; \boldsymbol\eta) + \sum_{j\in S} P_{n,\lambda}(|\beta_j|)
$$  
satisfying 
$$
||\hat{\boldsymbol{\beta}}_S - \boldsymbol\beta_{S0}||= O_p(a_n + \sqrt{s}P_{n,\lambda}'(d_n)), ~~~
||\boldsymbol\eta - \boldsymbol\eta_{0}|| = O_p(c_n).
$$
In addition, for any given $\epsilon>0$, the local minimizer $(\hat{\boldsymbol{\beta}}, \hat{\boldsymbol{\eta}})$ 
is strict with probability at least $1-\epsilon$ for sufficiently large $n$.
\label{THM:consistency_high}
\end{theorem}

We have assumed the true support $S$ to be known in the previous theorem, which is not the practical situation.
So, in the next theorem, for variable selection consistency, we will show that the true $S$ can be recovered from the data with probability tending to one.
This is equivalent to show that the local minimizer of $Q_{n,\lambda}$ restricted to $\mathcal{B}\times \mathbb{R}^d$, as obtained in the previous theorem,
is also a local minimizer on $\mathbb{R}^{p+d}$. 
To this end, we need further assumptions of the nature of the loss function at the local minimum obtained in the above theorem. 

\bigskip
\noindent
\textbf{Assumptions on the loss (L2):}\\
For the local minimizer $(\hat{\boldsymbol{\beta}}_S, \hat{\boldsymbol{\eta}})$ obtained in Theorem \ref{THM:consistency_high}, 
there exists a neighborhood $\mathcal{H} \subset \mathbb{R}^{p+d}$ of $(\hat{\boldsymbol{\beta}}_S^T, 0, \hat{\boldsymbol{\eta}}^T)^T$
such that, with probability tending to one,  we have 
$$
L_n(T\boldsymbol\beta,\boldsymbol\eta) -L_n(\boldsymbol\beta, \boldsymbol\eta) < \sum_{j\notin S}P_{n,\lambda}(|\beta_j|),
$$
for all $\boldsymbol\beta=(\boldsymbol\beta_S^T, \boldsymbol\beta_N^T)^T$ with $(\boldsymbol\beta^T, \boldsymbol\eta^T) \in \mathcal{H}$ and 
$\boldsymbol\beta_N\neq 0$.
Here, $T\boldsymbol\beta$ denote the projection of $\boldsymbol\beta$  onto the space generated by $S$, i.e., 
$T\boldsymbol\beta=(\beta_1', \ldots, \beta_p')^T$ with $\beta_j'=\beta_j I(j\in S)$.

\begin{theorem}[Variable selection optimality]
Under Assumptions (P), (L1) and (L2), we have the followings: 
\begin{itemize} 
\item[(i)] $(\hat{\boldsymbol{\beta}}_S, 0, \hat{\boldsymbol{\eta}})$ obtained in Theorem \ref{THM:consistency_high} 
is a local minimizer in $\mathbb{R}^{p+d}$ of  the general objective function 
$Q_{n,\lambda}(\boldsymbol\beta, \boldsymbol\eta)$ in (\ref{EQ:penal_GenObj}), with probability tending to one.

\item[(ii)] For any given $\epsilon >0$, the local minimizer $(\hat{\boldsymbol{\beta}}_S, 0, \hat{\boldsymbol{\eta}})$ is strict with probability at least $1-\epsilon$ 
for all sufficiently large $n$.
\end{itemize}
\label{THM:variable_high}
\end{theorem}

\subsection{The Linear Mixed-Effect Models}\label{SEC:Theory_LMM}

We will now look back to the linear mixed-effects model (\ref{EQ:model1}) 
and the corresponding penalized likelihood estimation minimizing (\ref{EQ:penal_log-likelihood}).
We will verify the general conditions of the two previous subsections for the corresponding likelihood loss given by (\ref{EQ:log-likelihood})
and present simplified results for the linear mixed model set-up.

First let us consider the low-dimensional set-up of Section \ref{SEC:Theory_fixed} with $p \leq n$
and re-label the observations $\{y_{ij}, \boldsymbol{X}_{ij}\}_{j=1,\ldots, n_i; i=1,\ldots,I}$ as $\{y_k, \boldsymbol{X}_k\}_{k=1,\ldots, n}$.
Let $D_k$ denotes the cluster indicator corresponding to the $k$-th observation in the re-labeled series with $D$ being the underlying random variable. 
Let us assume that $X$ is a stochastic variable with $Z\subset X$ , that the observations $\boldsymbol{V}_k=(y_k, \boldsymbol{X}_k, D_k)$, 
$k=1,\ldots, n$, are $n$ independent and identically distributed realizations of variables $(Y,\boldsymbol{X},D)$, 
and that Assumptions (M1) and (M3) hold for any regular distribution of the covariates.
Further, a straightforward but lengthy calculation shows that Assumption (M2) also holds for the linear mixed effect model (\ref{EQ:model1})
whenever $E(\boldsymbol{X}^t\boldsymbol{X})$ is finite and positive definite under the true distribution. 
Then, we have the asymptotic properties of the resulting penalized estimators from Theorems \ref{THM:fixed_consistency} and \ref{THM:fixed_oracle}
which is combined in the following proposition.

\begin{proposition}
Consider the set-up of the linear mixed-effects model (\ref{EQ:model1}) with stochastic covariates $X$ with $E(\boldsymbol{X}^t\boldsymbol{X})$ 
being finite and positive definite under the true distribution, having parameters $(\boldsymbol\beta_0, \boldsymbol\eta_0)$ and 
$(y_k\boldsymbol{X}_k,D_k)$ being i.i.d.. Assume the fixed low-dimensional parameter space with $p\leq n$ and 
$P_{n,\lambda}(\cdot) = n p_{\lambda_n}(\cdot)$ in the objective function (\ref{EQ:penal_log-likelihood}).
Then, we have the following:
\begin{enumerate} 
\item Under Assumption (PA1) on the penalty, there exists a local minimizer 
$(\hat{\boldsymbol{\beta}}, \hat{\boldsymbol{\eta}}) = ((\hat{\boldsymbol{\beta}}^{(1)T},\hat{\boldsymbol{\beta}}^{(2)T})^T, \hat{\boldsymbol{\eta}})$
of $Q_{n,\lambda}(\boldsymbol\beta,\boldsymbol\eta)$ in (\ref{EQ:penal_log-likelihood}) which satisfies the optimality properties in (\ref{EQ:results_3.1}).

\item Under Assumptions (PA2) with $\lambda_n\rightarrow 0$ and $\sqrt{n}\lambda_n \rightarrow \infty$ as $n\rightarrow\infty$, 
we have $\hat{\boldsymbol{\beta}}^{(2)} = 0$ with probability tending to one and the asymptotic distributions of $\hat{\boldsymbol{\beta}}^{(1)}$ and $\hat{\boldsymbol{\eta}}$ 
are given by (\ref{EQ:Asymp_beta_fixed}) and (\ref{EQ:Asymp_eta_fixed}) respectively, where we now have 
$$
\boldsymbol{I}_{1}(\boldsymbol\beta_{0}^{(1)}, \boldsymbol\eta_0) = \sum_{i=1}^{dim(\boldsymbol\beta_{0}^{(1)})} 
E\left[\boldsymbol{X}_i^T\boldsymbol{V}_i(\boldsymbol\eta_0)^{-1}\boldsymbol{X}_i\right]
$$ 
and $\boldsymbol{I}_{2}(\boldsymbol\beta_{0}^{(1)}, \boldsymbol\eta_0)$ can also be derived explicitly depending 
on the assumed structure of $\boldsymbol\Psi_{\boldsymbol\theta}$.
\end{enumerate}
\label{PROP:fixed_Lmm}
\end{proposition}

We have already noted that the $L_1$ penalized estimator of $\boldsymbol\beta$ is $\sqrt{n}$-consistence if we choose $\lambda_n=O_p(n^{-1/2})$,
which cannot be simultaneously satisfied with the second condition $\sqrt{n}\lambda_n \rightarrow \infty$ required for the oracle property.
Therefore, the usual LASSO with $L_1$ penalty cannot generate estimators which is simultaneously $\sqrt{n}$-consistent and also satisfy the oracle property.
The SCAD penalty, on the other hand, can generate estimators satisfying both the $\sqrt{n}$-consistency and oracle property for any suitably 
chosen regularization sequence $\{\lambda_n\}$, since only $\lambda_n\rightarrow 0$ is enough to ensure their consistency.

Next we will consider the high-dimensional set-up of $p>n$ as in Section \ref{SEC:Theory_high} and
present some simplified conditions for the linear mixed model (\ref{EQ:model1}) which in turn 
will imply the general Assumptions (L1) and (L2). 
For this set-up, we again assume that the observations $\boldsymbol{V}_k=(y_k, \boldsymbol{X}_k, D_k)$ are independent and identically distributed
for $k=1,\ldots, n$ and define $g(\boldsymbol{V}_k;\boldsymbol\beta,\boldsymbol\eta)= \sum_{i=1}^II(D_k=i)a_{ij_k}$,
where $j_k$ denotes the index $j$ of the $k$-th (relabeled) observation in the original labeling 
and $a_{ij}$ denotes the $j$-th element of the vector 
$\boldsymbol{a}_i=(\boldsymbol{Y}_i-\boldsymbol{X}_i^T\boldsymbol\beta)\boldsymbol{V}_i(\boldsymbol\eta)^{-1}$.
Now let us assume the followings for the linear mixed model set-up (\ref{EQ:model1}):
\begin{enumerate} 
\item[(A1)] There exists constants $b_1,b_2>0$ and $r_1, r_2 > 0$ satisfying, for all $t>0$,
$$
P(|g(Y, \boldsymbol{X}^T\boldsymbol\beta,\boldsymbol\eta)|>t) \leq \exp(-(t/b_1)^{r_1}),~~
\max_{l\leq p}P(|X^{(l)}|>t) \leq \exp(-(t/b_2)^{r_2}),
$$
where $X^{(l)}$ denotes the $l$-th coordinate of the covariate vector $\boldsymbol{X}$.

\item[(A2)] $\min_{j\in S} Var(g(Y, \boldsymbol{X}^T\boldsymbol\beta,\boldsymbol\eta)X^{(j)})$ is bounded away from zero for all $j=1,\ldots, p$.

\item[(A3)] $Var(X^{(j)})$ is bounded away from zero and $\infty$ uniformly in $j=1,\ldots,p$.


\item[(A4)] The eigenvalues of the matrices $\boldsymbol{I}_{1}(\boldsymbol\beta_{0}^{(1)}, \boldsymbol\eta_0)$ 
and $\boldsymbol{I}_{2}(\boldsymbol\beta_{0}^{(1)}, \boldsymbol\eta_0)$, defined in Proposition \ref{PROP:fixed_Lmm}, 
are bounded away from both zero and $\infty$.
\end{enumerate}
These assumptions are motivated from Assumptions 4.2-4.5 of Fan and Liao (2014) and can be shown to hold for the linear mixed model
with suitably chosen covariate distribution and mixed effects structure. 
Further, following Assumption 4.6 of Fan and Liao (2014), 
we assume the following additional condition on the penalty function $P_{n,\lambda}(\cdot)$ under the above mentioned set-up.

\bigskip
\noindent
\textbf{Assumptions on the penalty (P*):}
\begin{itemize}
\item[(i)] $P_{n,\lambda}'(d_n) = o(1/\sqrt{ns})$, $P_{n,\lambda}'(d_n)s^2 = O(1)$, $s\sqrt{\log p /n}=o(d_n)$ and
$$
sP_{n,\lambda}'(d_n) +s\sqrt{\log p/n} + s^3 \log s /n = o(P_{n,\lambda}'(0+)).
$$

\item[(ii)] $\displaystyle\sup_{||\boldsymbol\beta-\boldsymbol\beta_{S0}||\leq d_n/4} \zeta(\boldsymbol\beta) = o(1/\sqrt{s\log p}).$

\item[(iii)] $\displaystyle\max_{j\notin S}||\boldsymbol{X}_S^T\boldsymbol{V}_S^{-1}\boldsymbol{X}^{(j)}||\sqrt{\log s/n} = o(P_n(0+))$,
where $\boldsymbol{X}_S$ denotes the covariates corresponding to $\boldsymbol\beta_S$ and 
$\boldsymbol{V}_S$ being the summation of the associated $\boldsymbol{V}_i$ matrices. 
\end{itemize}
Based on these assumptions, one can easily show that the required assumptions (L1) and (L2) of Section \ref{SEC:Theory_high} hold 
as presented in the following lemma. Then, a direct application of Theorems \ref{THM:consistency_high} and \ref{THM:variable_high}
yields the corresponding asymptotic properties of the penalized estimators under the linear mixed model set-up, 
which in presented in the next proposition. The proofs are straightforward albeit lengthy and hence omitted for brevity.

\begin{lemma}
Under the above mentioned set-up of the linear mixed model with high-dimensionality, 
\begin{enumerate}
\item Assumptions (A1)--(A4) imply Assumption (L1) with $a_n=\sqrt{s\log p /n}$ and $c_n=1/\sqrt{n}$.
\item Assumptions (A1)--(A4) together with (P*) imply Assumption (L2).
\end{enumerate}
\end{lemma}

\bigskip
\begin{proposition}
Consider the set-up of the linear mixed model (\ref{EQ:model1}) with high-dimensionality as in Section \ref{SEC:Theory_high} 
such that $s^3 \log p = o(n)$. 
Assume that the observations $\boldsymbol{V}_k=(y_k, \boldsymbol{X}_k, D_k)$, $k=1,\ldots,n$, are i.i.d. and satisfy Assumptions (A1)--(A4) 
and the penalty function satisfies Assumptions (P) and (P*).
Then, there exists a local minimizer 
$(\hat{\boldsymbol{\beta}}, \hat{\boldsymbol{\eta}}) = ((\hat{\boldsymbol{\beta}}^{(1)T},\hat{\boldsymbol{\beta}}^{(2)T})^T, \hat{\boldsymbol{\eta}})$
of $Q_{n,\lambda}(\boldsymbol\beta,\boldsymbol\eta)$ in (\ref{EQ:penal_log-likelihood}) that satisfies 
\begin{enumerate} 
\item $\displaystyle\lim_{n\rightarrow\infty} P(\hat{\boldsymbol{\beta}}^{(2)} =0)=1$. 
In addition, the local minimizer is strict with probability arbitrarily close to one for all sufficiently large $n$.
 
\item Assuming $\hat{S} = \{j\leq p : \hat{{\beta}}_j \neq 0\}$ denotes the estimated active set, $\displaystyle\lim_{n\rightarrow\infty} P(\hat{S}=S)=1$.

\item For any unit vector $\boldsymbol\alpha\in\mathbb{R}^s$, 
$$
\sqrt{n}\boldsymbol\alpha^t \boldsymbol{I}_{1}(\boldsymbol\beta_{0}^{(1)}, \boldsymbol\eta_0)^{1/2}(\hat{\boldsymbol{\beta}}^{(1)}-\boldsymbol\beta_0^{(1)}) 
\mathop\rightarrow^\mathcal{D} N(0,1),
~~~~
\sqrt{n}(\hat{\boldsymbol{\eta}}-\boldsymbol\eta_0) \mathop\rightarrow^\mathcal{D} N_d(0,\boldsymbol{I}_{2}(\boldsymbol\beta_{0}^{(1)},\boldsymbol\eta_0)^{-1}).
$$
\end{enumerate}
\label{PROP:high_Lmm}
\end{proposition}

\bigskip
Note that Assumption (P*) imposes restrictions on the required lower bound on the signal $d_n$ 
in terms of the number of important fixed effect variables ($s$) and the penalty function used. 
This Assumption (P*) can be seen to hold for the SCAD penalty whenever  the signal $d_n$ is strong enough 
and $s$ is small compared to the total sample size $n$ such that 
$s\sqrt{\log p/n} + s^3\log s /n \ll \lambda_n \ll d_n$.
These types of assumptions are quite common in the high-dimensional set-up and are required mainly to achieve the variable selection consistency.
See Remarks 4.3 and 4.4 of Fan and Liao (2014) for some related discussions on similar assumptions in the context of linear regression.


\section{Computational Algorithm}\label{SEC:computation}

Since the minimization problem in finding the MPLE is a non-convex optimization problem, standard
approaches fail and we need some suitable iterative algorithm to obtain the MPLEs efficiently.
We will follow the unified approach provided in Fan and Li (2001) which uses some local quadratic approximation 
to the objective function and then use the iterative Newton-Raphson algorithm. 
However, to achieve greater computational efficiency in the cases with large $p$
we will combine it with a version of the co-ordinate descent algorithm.
Consider the general optimization problem with objective function given by (\ref{EQ:penal_GenObj}).
We will first present the quadratic approximation for this general objective function 
and then describe the coordinate descent algorithm to obtain its minimizer. 

\subsection{Quadratic Approximation of the Objective function}\label{SEC:Quad_approx}

In order to get a quadratic approximation of the general objective function $Q_{n,\lambda}(\boldsymbol\beta,\boldsymbol\eta)$ in (\ref{EQ:penal_GenObj}),
we need the same for both the loss function $L_n(\boldsymbol\beta,\boldsymbol\eta)$ and the penalty $p_\lambda(\cdot)$.
Since the first one is generally a function of log-likelihood, it is quite easy to get a quadratic approximation of this term both with respect to $\beta$ and $\eta$. 
We have to just assume that the loss function $L_n(\boldsymbol\beta,\boldsymbol\eta)$ is smooth with respect to both parameters, 
having continuous second order partial derivatives, which is generally true for most common statistical models. 
Then, using the Taylor series approach, we have the following quadratic approximation with respect to $\boldsymbol\beta$ and $\boldsymbol\eta$:
\begin{eqnarray}
L_n(\boldsymbol\beta,\boldsymbol\eta) 
&\approx& L_n(\boldsymbol\beta_0, \boldsymbol\eta_0) + \nabla_\beta L_n(\boldsymbol\beta_0,\boldsymbol\eta_0)^T (\boldsymbol\beta - \boldsymbol\beta_0) 
+ \nabla_\eta L_n(\boldsymbol\beta_0,\boldsymbol\eta_0)^T (\boldsymbol\eta - \boldsymbol\eta_0)
\nonumber\\&& ~
+ \frac{1}{2}(\boldsymbol\beta - \boldsymbol\beta_0)^T\nabla_\beta^2 L_n(\boldsymbol\beta_0,\boldsymbol\eta_0) (\boldsymbol\beta - \boldsymbol\beta_0) 
+ \frac{1}{2}(\boldsymbol\eta - \boldsymbol\eta_0)^T\nabla_\eta^2 L_n(\boldsymbol\beta_0,\boldsymbol\eta_0) (\boldsymbol\eta - \boldsymbol\eta_0)
\nonumber\\&& ~
+ (\boldsymbol\beta - \boldsymbol\beta_0)^T\nabla_{\beta\eta} L_n(\boldsymbol\beta_0,\boldsymbol\eta_0) (\boldsymbol\eta - \boldsymbol\eta_0),
\label{EQ:Quad_loss}
\end{eqnarray}
where $\nabla_\beta L_n(\boldsymbol\beta_0,\boldsymbol\eta_0)$ and $\nabla_\eta L_n(\boldsymbol\beta_0,\boldsymbol\eta_0)$ 
are the first order partial derivatives of $L_n(\boldsymbol\beta_0,\boldsymbol\eta_0)$ with respect to $\boldsymbol\beta$ and $\boldsymbol\eta$ respectively, 
$\nabla_\beta^2 L_n(\boldsymbol\beta_0,\boldsymbol\eta_0)$ and $\nabla_\eta^2 L_n(\boldsymbol\beta_0,\boldsymbol\eta_0)$ 
are corresponding second order partial derivatives and $\nabla_{\beta\eta} L_n(\boldsymbol\beta_0,\boldsymbol\eta_0)$ 
is the second order cross-partial derivatives with respect to $\boldsymbol\beta$ and $\boldsymbol\eta$ sequentially.

However, the general non-concave penalty functions described in Section \ref{SEC:penalty} do not generally posses everywhere continuous derivatives.
In particular, the $L_1$ penalty, hard threshold penalty and even the SCAD penalty function do not have continuous second order derivatives at the origin
and so we cannot use the above mentioned Taylor series approach to get the quadratic approximation for the penalty functions.
So, we will follow the local quadratic approximation of the penalties as described in Fan and Li (2001). 
Note that the penalty term involves only the parameter $\boldsymbol\beta$ and not $\boldsymbol\eta$.
Given an initial value $\boldsymbol\beta_0$ close to the actual minimizer, if its $j$-th component $\beta_{0j}$ is not very close to zero
we can use the local approximation
\begin{equation}
[p_\lambda(|\beta_j|)]' = p_\lambda'(|\beta_j|) sgn(\beta_j) \approx \left\{\frac{p_\lambda'(|\beta_j|)}{|\beta_j|}\right\}\beta_j
\end{equation}      
and set $\hat{\boldsymbol{\beta}}_j=0$ if $\beta_{0j}$ is very close to zero. 
Combining it with the Taylor series expansion, we get the quadratic (local) approximation for the penalty function as 
\begin{eqnarray}
p_\lambda(|\beta_j|) &\approx& p_\lambda(|\beta_{0j}|) + \frac{1}{2}\left\{\frac{p_\lambda'(|\beta_j|)}{|\beta_j|}\right\}(\beta_{j}^2 - \beta_{0j}^2),
\label{EQ:Quad_penalty}
\end{eqnarray}  
for $\beta_j \approx \beta_{0j}$.

Therefore the general objective function can be locally approximated by a quadratic function and 
the Newton-Raphson method can be used to minimize it if the number of parameters is small.
In particular, this minimization problem leads to the following iterative solution
\begin{eqnarray}
\boldsymbol\beta_1 &=& \boldsymbol\beta_0  - \left[\nabla_\beta^2 L_n(\boldsymbol\beta_0,\boldsymbol\eta_0) 
+ \boldsymbol\Sigma_\lambda(\boldsymbol\beta_0)\right]^{-1}
\left\{\nabla_\beta L_n(\boldsymbol\beta_0,\boldsymbol\eta_0) + \boldsymbol\Sigma_\lambda(\boldsymbol\beta_0)\boldsymbol\beta_0 \right\},
\label{EQ:Sol_beta}\\
\boldsymbol\eta_1 &=& \boldsymbol\eta_0  - \left[\nabla_\eta^2 L_n(\boldsymbol\beta_0,\boldsymbol\eta_0)\right]^{-1}
\left\{\nabla_\beta L_n(\boldsymbol\beta_0,\boldsymbol\eta_0) \right\},
\label{EQ:Sol_eta}
\end{eqnarray}
where $\boldsymbol\Sigma_\lambda(\boldsymbol\beta)=Diag\left\{\frac{p_\lambda'(|\beta_j|)}{|\beta_j|}\right\}_{j=1,\ldots,p}$.
We can iterate sequentially within $\beta$ and $\eta$ until convergence to obtain the minimizer of the general objective function in (\ref{EQ:penal_GenObj}).

This algorithm has been checked to work when the number of parameters is less than the sample size 
and it converges quite efficiently for different penalty functions.
However, for the high-dimensional set-up where the number of parameter is larger than the sample size, 
the above iteration scheme fails at the step of updating the $\boldsymbol\beta$ using the inverse of a large matrix 
and the task becomes computationally unstable and inefficient.
So wee need to modify our algorithm by using a suitable co-ordinate descent algorithm for the update of $\boldsymbol\beta$ in (\ref{EQ:Sol_beta})
as described in the next subsection.

\subsection{Coordinate Gradient Descent Algorithm for High-dimensional set-up}\label{SEC:QDA}

There exist several proposals based on different versions of the coordinate-descent approach in the high-dimensional situation 
under the regression set-up without $\boldsymbol\eta$. 
These are mainly based on coordinate-wise optimization for the high-dimensional vector $\boldsymbol\beta$ following the idea of Tseng and Yun (2009)
and was applied in different high-dimensional regressions, for example, penalized least square with $L_q$ penalty (Fu, 1998; Daubechies, Defrise and De Mol, 2004),
penalized estimation of the precision matrix (Friedman et al., 2007), ordinary linear Lasso (Wu and Lang, 2008), grouped Lasso (Meier, van de Geer and Bühlmann, 2008),
 Lasso for generalized linear models (Friedman, Hastie and Tibshirani, 2010), nonconcave penalized GLM (Fan and Lv, 2011) and many more.
Schelldorfer et al.~(2011) used such a coordinate gradient descent (CGD) algorithm for the high-dimensional linear mixed model with $L_1$ penalty
incorporating the optimization of additional variance parameters $\boldsymbol\eta$ as well.
Here, we will follow their CGD approach for solving our optimization problem with general non-concave penalty in the high-dimensional situation.   
For minimizing the general objective function in (\ref{EQ:penal_GenObj}), the CGD algorithm works as follows: 

\bigskip
\noindent
\textbf{CGD Algorithm:}
\begin{enumerate}
	\item Start with a initial value $\boldsymbol\beta^0$ and $\boldsymbol\eta^0$. 
	\item For $j=1,2, \ldots, p$,
	\begin{enumerate}
	\item Approximate the second order derivative $\frac{\partial^2}{\partial\beta_j^2}Q_{n,\lambda}(\boldsymbol\beta, \boldsymbol\eta)$ by (Tseng and Yun, 2009)
	$$
	h_j = \min\{\max\{\mathcal{I}_{jj}, c_{min}\}, c_{max}\},
	$$
	for some suitable constants $c_{min}$ and $c_{max}$. Here, $\mathcal{I}_{jj}$ denotes the $j$-th diagonal element of 
	the Fisher information matrix $\boldsymbol{I}(\boldsymbol\beta,\boldsymbol\eta)$ of $\boldsymbol\beta$  
	(Schelldorfer et al., 2011, suggested the choice $c_{min}=10^{-6}$ and $c_{max}=10^{8}$).
	\item Calculate the gradient direction $d_j$ by minimizing, with respect to $d\in\mathbb{R}$, the following
	$$
	L_n(\boldsymbol\beta,\boldsymbol\eta) + \frac{\partial}{\partial\beta_j}L_{n}(\boldsymbol\beta, \boldsymbol\eta) d 
	+ \frac{1}{2}d^2h_j + P_{n,\lambda}(|\beta_j+d|).
	$$
	\item Choose a step-size $\alpha_j>0$ such that 
	$Q_{n,\lambda}(\boldsymbol\beta+\alpha_j d_j \boldsymbol{e}_j, \boldsymbol\eta)<Q_{n,\lambda}(\boldsymbol\beta, \boldsymbol\eta)$,
	where $\boldsymbol{e}_j$ denotes the $j$-th unit vector. If we can find such an $\alpha_j>0$ then update the $j$-th component of $\boldsymbol\beta$ as
	$$
	\beta_j \leftarrow \beta_j + \alpha_j d_j.
	$$
	As in Schelldorfer et al.~(2011), this $\alpha_j$ can be chosen by the Armijo rule described in Remark \ref{REM:Armijo} below.  
	\end{enumerate}
	\item For $j=1,2,\ldots,d$,
	update $\eta_j$ by the minimizer of $L_n(\boldsymbol\beta,\boldsymbol\eta)$ with respect to $\eta_j$ 
	with updated $\boldsymbol\beta$ and $\boldsymbol\eta$ from the previous steps.
	
	\item Repeat Steps 2 and 3 until convergence.
\end{enumerate}

\bigskip
\begin{remark}[Armijo Rule]
\label{REM:Armijo}
Following Schelldorfer et al.~(2011), the Armijo rule for obtaining $\alpha_j$ can be defined as follows: 
Start with an initial value $\alpha_j^{(0)}$ and define $\alpha_j$ as the largest element of $\{\alpha_j^{(0)}\delta^r\}_{r=0, 1, 2 \ldots}$
that satisfies 
$$
Q_{n,\lambda}(\boldsymbol\beta+\alpha_j d_j \boldsymbol{e}_j, \boldsymbol\eta) \leq Q_{n,\lambda}(\boldsymbol\beta, \boldsymbol\eta) + \alpha_j \rho \Delta_j, 
$$
where $\Delta_j = \frac{\partial}{\partial\beta_j}L_{n}(\boldsymbol\beta, \boldsymbol\eta) d_j 
+ \gamma d_j^2h_j + P_{n,\lambda}(|\beta_j+d|)-P_{n,\lambda}(|\beta_j|)$.
Suggested choices for the constants are (Bertsekas, 1999) $\delta=0.1$, $\rho=0.001$, $\gamma=0$ and $\alpha_j^{(0)}=1$ for all $j$.
\end{remark}

It is to be noted that, the above algorithm is crucially dependent on the starting value used and 
also may not converge to the global optimum due to the non-convexity of the objective function.
However, convergence of the algorithm to a local optimum is certain as shown in Theorem 3 of Schelldorfer et al.~(2011).
Also, regarding the choice of initial values, we can choose an optimum ordinary Lasso solution for $\boldsymbol\beta$ ignoring the mixed-effect structure
which ensures that we are at least as good as the ordinary Lasso objective function. 
Throughout the present paper, we have used the 10-fold cross validated Lasso estimate of $\boldsymbol\beta$ as the initial value  
$\boldsymbol\beta^0$ in all illustrations.
The initial value $\boldsymbol\eta^0$ for the variance parameter $\boldsymbol\eta$ depends on the assumed variance structure and 
can be obtained by suitable Gauss-Seidel iteration based on the usual maximum likelihood principal.

We can simplify Steps 2(b) and 2(c) further depending on the structure of the penalty function used. 
Such simplified calculations for the $L_1$-penalty can be found in Appendix C of Schelldorfer et al.~(2011).
We will briefly present the simplified calculations for the SCAD penalty in Remark \ref{REM:Algo_SCAD} below.

\bigskip
\begin{remark}[Simplification in Algorithm for SCAD penalty]
\label{REM:Algo_SCAD}
For the SCAD penalty, the quantity $d_j$ defined in Step 2(b) of the CGD algorithm can be calculated analytically. 
Whenever $\beta_j$ is not subject to penalization it has the form
$$
d_j = - \frac{1}{h_j}\frac{\partial}{\partial\beta_j}L_{n}(\boldsymbol\beta, \boldsymbol\eta), 
$$
whereas if $\beta_j$ is subject to penalization through the SCAD penalty with regularization parameters $\lambda$ and $a$ 
then the solution $d_j$ is given by 
\begin{equation}
d_j = \left\{ 
\begin{array}{ll}
- \frac{1}{h_j}\left[\lambda+\frac{\partial}{\partial\beta_j}L_{n}(\boldsymbol\beta, \boldsymbol\eta)\right]  	
& \mbox{ if }   \beta_j h_j - \frac{\partial}{\partial\beta_j}L_{n}(\boldsymbol\beta, \boldsymbol\eta)\leq \lambda (h_j+1),\\
- \frac{1}{h_j}\frac{\partial}{\partial\beta_j}L_{n}(\boldsymbol\beta, \boldsymbol\eta)  									
& \mbox{ if }  \beta_j h_j - \frac{\partial}{\partial\beta_j}L_{n}(\boldsymbol\beta, \boldsymbol\eta) > \lambda (a h_jj),\\
- \frac{1}{(a-1)h_j-1}\left[(a-1)\frac{\partial}{\partial\beta_j}L_{n}(\boldsymbol\beta, \boldsymbol\eta) - (a\lambda-\beta_j)\right]			& \mbox{ otherwise. }  
\end{array}\right.
\label{EQ:SCAD_dj}
\end{equation}
Further, if $h_j=\mathcal{I}_{jj}$, i.e., no truncation is used, 
we can also get an analytical solution for the update of $\beta_j$ is Step 2(c) of CGD algorithm 
based on the solution (\ref{EQ:SCAD_sol}) of the SCAD penalized likelihood.
For the linear mixed effect model, it is given based on $h_j=\boldsymbol{x}_j^T\boldsymbol{V}^{-1}\boldsymbol{x}_j$ and 
$\hat\gamma((y-\tilde{y})\boldsymbol{V}^{-1}\boldsymbol{x}_j/h_j)$,
where $\tilde{y}$ is the (marginal) predicted value of $y$ based on all the fixed effects except the $j$-th one.
\end{remark}

\subsection{Choice of the regularization parameters}\label{SEC:Choice_par}

The next computational challenge is the selection of the regularization parameter $\lambda$,
which is very important to get the optimal performance of the proposal for any penalty.
Fan and Li (2001) considered cross-validation for their linear regression model, 
but the objective function to be used in cross-validation is not quite clear for the mixed model. 
Instead  Schelldorfer et al.~(2011) proposed to use BIC for selection of $\lambda$ which can be adopted for our set-up also.

For the mixed effect model (\ref{EQ:model1}),  the BIC can be defined as (Schelldorfer et al., 2011)
\begin{equation}
{\rm BIC}_\lambda = -2 l_n(\hat{\boldsymbol{\beta}},\hat{\boldsymbol{\eta}}) + (\log n) \hat{df}_\lambda,
\label{EQ:BIC}
\end{equation}
where the degrees of freedom for the mixed model can be estimated as $|\{1\leq j \leq p : \hat{{\beta}}_j \neq 0 \}|+dim(\eta)$.
The factor $|\{1\leq j \leq p : \hat{{\beta}}_j \neq 0 \}|$ is actually the expected degrees of freedom in the ordinary linear Lasso (Zou, Hastie and Tibshirani, 2007) 
and the dimension of the additional parameters has been added to get the corresponding estimate for the linear mixed model set-up.
This process gives very good results in choosing optimal $\lambda$ for all kind of penalties as to be seen in the next section.

For penalties like SCAD, where we have one additional parameter $a$, this can also be chosen by the above minimum BIC approach.
However, for the SCAD penalty, Fan and Li (2001) proposed $a=3.7$ to be the optimal choice in terms of Bayes risk 
and shown to provide equivalent results compared to the value chosen by general cross-validation in the context of the linear regression model.
So, in the present paper, we will also fix $a=3.7$ in all illustrations with SCAD penalty.

\section{Numerical Illustrations}\label{SEC:numerical}
 
\subsection{Simulation Study}\label{SEC:Simulation}

In this section, we will present the finite sample performance of the proposed method with the SCAD penalty through 
a suitable simulation study and compare the results with the standard $L_1$ penalty. 
As the main objective of the paper is to focus on the selection of fixed effects and their estimation,
these particular issues are examined for several linear mixed-effect  models with different true parameters 
and design matrices for both the $L_1$ and SCAD penalization.
Considering the similarity of results, we present only some selected cases for both the low and high-dimensional set-ups.
The regularization tuning parameter $\lambda$ is chosen by minimizing the BIC for each of these simulations separately
whereas the parameter $a$ in the definition of the SCAD penalty has been kept fixed at $a=3.7$ for all.

In particular, we  present the results for a linear mixed model set-up with number of groups $I=25$ and $n_i=6$ observations per group
leading to a sample size of $n=150$. We will choose several numbers of fixed effects as $p=10,50$ (low-dimensional set-up) 
and also $p=300,500$ (high-dimensional set-up). However, in all the cases we will keep the size of the active set to be $s=5$
with the true value of the fixed effect coefficient $\boldsymbol\beta$ being $\boldsymbol\beta_0=(1,2,4,3,3,0, \ldots, 0)^T$.
The number of random effects are chosen as $q=2$ with the random effect coefficients being normally distributed with mean $0$ and covariance matrix 
$\boldsymbol\Psi_\theta=\theta^2 I_2$, where $I_2$ is the identity matrix of dimension 2. The true values of the  variance components are assumed to be 
$\theta^2=0.56$ and $\sigma^2=0.25$ (error variance). 
The design matrix $\boldsymbol{X}$ is chosen such that the first column is 1 yielding the intercept and the next $(p-1)$ columns are chosen 
from a multivariate normal distribution with mean $0$ and a covaraince matrix having $(i,j)$-th element as $\rho^{|i-j|}$
for all $i,j=1,\ldots, p-1$. In our simulation, we have chosen $\rho=0$ giving the case of independent covariates and also $\rho=0.5$
generating the case of correlated covariates. 
The random effect covariates are chosen as the first $q=2$ columns of the fixed effects design matrix;
so we have one random intercept and one random slope in our simulation. 
Motivated from the findings of Schelldorfer et al.~(2011) for the $L_1$-penalty in high-dimension, 
we have also kept the first two covariates (which also appear in the random effect part) 
including the intercept term as non-penalized in the fixed-effects counterpart for all the cases.

The empirical mean, standard deviation (SD) and the mean squared error (MSE) of the parameter estimates over $100$ iterations has been reported in 
Tables \ref{TAB:Sim_L1} and \ref{TAB:Sim_SCAD} for the $L_1$ and the SCAD penalty for all our simulation set-ups. 
For the fixed-effects coefficients not in the true active set, we only report their average values as $\boldsymbol\beta_N$.
We have also reported the average value of the estimated active set size ($|S(\hat{\boldsymbol\beta})|$) and the number of true positives (TP) 
in the tables along with their SD over those 100 iterations,
and the same for the model prediction error (PE) obtained after adjustment for the random effects. 
These random effect components are predicted through the maximum a posteriori (MAP) approach of Schelldorfer et al.~(2011).
Several important observations on the properties of our proposed method can easily be made based on the results 
from Tables \ref{TAB:Sim_L1} and \ref{TAB:Sim_SCAD}, which include the following: 
\begin{itemize}
\item In terms of variable selection, both the SCAD and $L_1$ penalty based methods correctly identify all the true positives in all cases.
Further, SCAD based method generally chooses less false positives compared to that based on $L_1$ penalty based method, yielding a smaller active set; 
this improvement becomes more and more clear in high-dimensional set-ups and in cases with uncorrelated covariates. 
This clearly shows the usefulness of our proposed SCAD penalty in terms of variable selection with high-dimensionality.

\item In terms of model prediction and related error, both the $L_1$ and SCAD penalty based methods performs similarly. 
The observed PE is significantly small in all the cases.

\item Estimation of fixed effect coefficients are also quite competitive for both penalties, with the SCAD penalty 
providing slightly lesser bias and MSE in most cases. 
For the first two components which also involves some random effects, these have greater standard error for both penalties
while their bias is slightly less for the SCAD penalty in case with correlated covariates.
For all other components of $\boldsymbol\beta$, bias is almost negligible and standard errors are also quite low for both penalties,
with SCAD providing slightly improved results.

\item As expected from the theory of likelihood in mixed models, 
the estimates of the variance component parameters $\sigma^2$ and $\theta^2$ have a downward bias for both penalties.
However, the bias in $\sigma^2$ is quite small for both penalties and becomes even negligible in low-dimension for the SCAD penalty.
The downward bias and SD of the estimates of $\theta^2$ is higher as expected and are similar for both penalties, 
although SCAD again provide some improvements for very high-dimensional set-up with correlated covariates ($p=500, \rho=0.5$).
\end{itemize}

It is clear from this and from other simulation studies not reported here, that 
if the focus is on selection and estimation of fixed effect components, 
the proposed SCAD penalty performs clearly better than the existing $L_1$ penalty in high-dimensional set-ups. 
These observations combined with the better theoretical properties as illustrated in Section \ref{SEC:Theory_LMM}
strongly promote our proposal of SCAD penalty in high-dimensional linear mixed effect models.

\subsection{A Real Data Application}\label{SEC:real_data}

Ottestad et al. (2012) investigated the effects of intake of oxidized and non-oxidized fish oil on inflammatory markers 
in a randomized study of 52 subjects. Inflammatory markers were measured at baseline and after three and seven weeks. 
In this example we will use the same data to investigate whether there are any associations 
between gene expressions measured at baseline and level of the inflammatory marker ICAM-1 throughout the study.

Although no associations were found between treatment and inflammatory markers, 
we decided to respect the design of the study by including fixed effects of treatment (3 groups), time and their interaction ``Treatment $\times$ Time", in addition to the gene expression measurements. 
From a vast set of genes, we initially selected $p^*=506$ genes having absolute correlation greater than or equal to $0.2$ with the response at any time point, 
so that the total number of fixed effects considered becomes $p=p^*+6 = 512$.
On the other hand, removing the missing observations in the response variable $y_{ij}$ for some $i$, $j$, 
we obtain $n=150$ observations, making it a high-dimensional selection problem. 
Further, due to the longitudinal structure of the data, we additionally considered random effect components in the model;
we included a random intercept ($b_I$) and a random slope corresponding to the $``Time"$ variable ($b_{Time}$) and assume that 
$(b_I, b_{Time})^T \sim N_2(0, Diag\{\hat{\theta}_I^2, ~\hat{\theta}_{Time}^2\})$. 
Then we applied the penalized maximum likelihood estimation
with the proposed SCAD penalty as well as the classical $L_1$ penalty; 
the regularization parameter $\lambda$ was chosen by minimizing the BIC as in the simulation study.
In Table \ref{TAB:real_beta}, we present the estimated fixed effect coefficients for all the genes that were selected by at least one of the methods,
and in Table \ref{TAB:real_eta} we present the estimated variance component parameters. 
We have also presented the estimated coefficients of the fixed effect variables under a simple linear regression model 
ignoring the mixed-effect structure in Table \ref{TAB:real_beta}; 
these are computed using a 10-fold cross validated penalized maximum likelihood estimation and 
are used as the initial estimates for the computation in mixed model set-up as before. 
To study the usefulness of these methods, in Table \ref{TAB:real_beta}, 
we have also marked the genes by $(*)$ which are known to be related to the inflammation and immune response  from biological mechanisms. 
Some genes which are not properly identified are marked as ``NA".

We can notice that the mixed models based on the SCAD penalty and the $L_1$ penalty, respectively, 
select about the same number of genes (29 for SCAD, 30 for L1). 
Furthermore, the active set becomes significantly smaller in the mixed model set-up compared 
to the ordinary linear regression models that ignore the dependence, 
implying the actual need for applying a mixed model for these data. 
We can also notice that when looking at the ten largest estimated $\beta$'s (in absolute value), 
six of them are known to be associated with inflammation when applying the SCAD penalty, 
while only four of the known genes are picked up by the $L_1$ penalty.

Looking at the estimated random effects (Table \ref{TAB:real_eta}), it is worth pointing out that the estimated random intercept variation 
is zero when the gene expressions are included in the model. 
It should also be mentioned that the error variance $\sigma^2$ is slightly reduced for the SCAD penalty as compared to the $L_1$ penalty.
 in these data.

\section{Discussions/Concluding Remarks}\label{SEC:Conclusion}

In this paper, we have discussed general non-concave penalty functions for penalized likelihood based 
parameter estimation and fixed effects selection in the linear mixed model. 
Asymptotic properties like consistency and oracle property of variable selection has been proved for 
the general non-convex loss under both  low and high-dimensional set-up.
Corresponding results for the linear mixed model set-up has been obtained as a particular case
and the usefulness of the SCAD penalty function has been illustrated through improved  
asymptotic properties and numerical performances over the classical $L_1$ penalty. 
This complements the only existing theory of high-dimensional mixed models using $L_1$ penalty.

For the low-dimensional set-up, we also provided the asymptotic distribution of the penalized estimators under general loss and penalty. 
However, for high-dimensional set-up, due to technical difficulties, the asymptotic distribution of the penalized estimators has been 
provided only for the linear mixed-effect model but with a general class of non-concave penalty functions. 
It will be an interesting future work to extend this to obtain asymptotic distribution for general loss under high-dimensionality,
but this will require more strict conditions. 

Although we have only considered the linear mixed model in the present paper, 	
the proposal can be extended to the case of generalized linear mixed models (GLMMs) also.
The general theory provided in Subsections \ref{SEC:Theory_fixed} and \ref{SEC:Theory_high} also cover the likelihood functions of the GLMMs 
and suitable simplifications as in Subsection \ref{SEC:Theory_LMM} can be obtained for them.
However, the challenge will be to obtain an efficient numerical optimization algorithm for GLMMs with high-dimensionality,
which we want to explore in a subsequent research paper.

The paper also opens the possibility of many future works in the high-dimensional mixed effect models. 
In particular, the hypothesis testing issue has not been considered till now under the high-dimensional mixed models,
although there are some recent attempts for high-dimensional regression models. 
This work provides a ground for extending them from regression to mixed model set-up
since we have first developed an asymptotic distribution for the parameter estimates in high-dimensional linear mixed models.
Further, studying the effect of violation of the assumptions like exogeneity, normality etc in the high-dimensional set-up 
and their possible remedies will also be very useful from a practical point-of view.
We hope to pursue some of these in the future.

\bigskip\noindent
\textbf{Acknowledgment:} 
The work is funded by the Norwegian Cancer Society, grant no.~5818504.
We also thanks Prof.~Stine Ulven from the department of Nutrition, University of Oslo, 
for providing the real dataset used in the paper and 
also for her help and guidance in biological interpretation of the results.



\appendix
\section{Appendix: Proofs}\label{APP:proof}

\subsection{Proof of Theorem \ref{THM:fixed_consistency}}

This proof is an extension of the proof of Theorem 1 of Fan and Li (2011).\\
Let us denote $\alpha_n=n^{-1/2} + a_n$ and fix any $\epsilon>0$. 
We will show that there exists a constant $C>0$ such that 
\begin{equation}
P\left(\sup_{\boldsymbol{u}=(\boldsymbol{u}_1^T, \boldsymbol{u}_2^T)^T \in \mathbb{R}^{p+q}: ||\boldsymbol{u}||=C} 
Q_{n,\lambda}(\boldsymbol\beta_0 + \alpha_n \boldsymbol{u}_1, \boldsymbol\eta_0+n^{-1/2}\boldsymbol{u}_2) > Q_{n,\lambda}(\boldsymbol\beta_0,\boldsymbol\eta_0)\right) \geq 1-\epsilon.	
\label{EQ:Toshow_3.1}
\end{equation}
It will then follows that there exist a local minimizer of $Q_{n,\lambda}(\boldsymbol\beta,\boldsymbol\eta)$ in the ball 
$\{\boldsymbol\beta_0 + a_n \boldsymbol{u}_1, \boldsymbol\eta_0+n^{-1/2}\boldsymbol{u}_2 : ||(\boldsymbol{u}_1^T, \boldsymbol{u}_2^T)^T||\leq C\}$
and hence that minimizer satisfies (\ref{EQ:results_3.1}) with probability at least $\epsilon$.

Now, using the property $p_{\lambda_n}(0)=0$, 
\begin{eqnarray}
D_n(\boldsymbol\beta,\boldsymbol\eta) &=& Q_{n,\lambda}(\boldsymbol\beta_0 + \alpha_n \boldsymbol{u}_1, \boldsymbol\eta_0+n^{-1/2}\boldsymbol{u}_2)  - Q_{n,\lambda}(\boldsymbol\beta_0,\boldsymbol\eta_0) \nonumber \\
&\geq& L_n(\boldsymbol\beta_0 + \alpha_n \boldsymbol{u}_1, \boldsymbol\eta_0+n^{-1/2}\boldsymbol{u}_2) - L_n(\boldsymbol\beta_0,\boldsymbol\eta_0)
+ \sum_{j=1}^s n\left[p_{\lambda_n}(\beta_{0j} + a_n u_{1j}) - p_{\lambda_n}(|\beta_{0j}|)\right]\nonumber \\
&=& \left\{\alpha_n\nabla_\beta L_n(\boldsymbol\beta_0,\boldsymbol\eta_0)^T\boldsymbol{u}_1 
+ \frac{1}{2} \boldsymbol{u}_1^T\boldsymbol{I}_{11}(\boldsymbol\beta_0,\boldsymbol\eta_0)\boldsymbol{u}_1 n \alpha_n^2(1+o_P(1))\right\}\nonumber \\
&&~~+ \left\{n^{-1/2}\nabla_\eta L_n(\boldsymbol\beta_0,\boldsymbol\eta_0)^T\boldsymbol{u}_2 
+ \frac{1}{2} \boldsymbol{u}_2^T\boldsymbol{I}_{22}(\boldsymbol\beta_0,\boldsymbol\eta_0)\boldsymbol{u}_2 (1+o_P(1))\right\}\nonumber\\
&&~~+ \boldsymbol{u}_1^T\boldsymbol{I}_{12}(\boldsymbol\beta_0,\boldsymbol\eta_0)\boldsymbol{u}_2 \alpha_n n^{-1/2}(1+o_P(1)) \nonumber\\
&&~~+\sum_{j=1}^s n\left[\alpha_n p_\lambda'(\beta_{0j})sgn(\beta_{0j})u_{1j} + \alpha_n^2 p_\lambda''(|\beta_{0j}|)u_{1j}^2(1+o_P(1))\right],\nonumber
\end{eqnarray}
using a standard Taylor series argument. 
Here, $\boldsymbol{I}(\boldsymbol\beta_0,\boldsymbol\eta_0) = \begin{pmatrix}
\boldsymbol{I}_{11}(\boldsymbol\beta_0,\boldsymbol\eta_0) & \boldsymbol{I}_{12}(\boldsymbol\beta_0,\boldsymbol\eta_0)\\
\boldsymbol{I}_{12}(\boldsymbol\beta_0,\boldsymbol\eta_0)^T & \boldsymbol{I}_{22}(\boldsymbol\beta_0,\boldsymbol\eta_0)
\end{pmatrix}$ 
denote the partition with $\boldsymbol{I}_{11}(\boldsymbol\beta_0,\boldsymbol\eta_0)$ being of dimension $p\times p$.
Now, since $n^{-1/2}\nabla_\beta L_n(\boldsymbol\beta_0,\boldsymbol\eta_0)=O_P(1)$, the first term in the first bracket 
is $O_p(n^{1/2}\alpha_n)=O_P(n\alpha_n^2)$ and hence is uniformly dominated by the  second term  within the same bracket in $||\boldsymbol{u}_1||=C_1$ 
for some suitably chosen large $C_1>0$.  
Similarly, since $n^{-1/2}\nabla_\eta L_n(\boldsymbol\beta_0,\boldsymbol\eta_0)=O_P(1)$, the first term in the second bracket
is $O_p(1)$ and hence is uniformly dominated by the corresponding second term  in $||\boldsymbol{u}_2||=C_2$ 
for suitable $C_2>0$.  Finally, the last term is bounded by
$$
\sqrt{s} \alpha_n n a_n ||\boldsymbol{u}_1|| + \alpha_n^2 n b_n ||\boldsymbol{u}_1||^2,
$$
which is again bounded by the second term in the last bracket uniformly in $||\boldsymbol{u}_1||=C_1$. 
Hence,  (\ref{EQ:Toshow_3.1}) holds for choosing $C> C_1 + C_2 >0$ and using Assumption (PA1) and the fact that $\boldsymbol{I}(\boldsymbol\beta, \boldsymbol\eta)$ is positive definite. 
This completes the proof.

\subsection{Proof of Theorem \ref{THM:fixed_oracle}}
This proof is an extension of the proof of Theorem 2 of Fan and Li (2011).\\
We will first present the following Lemma which is a generalization of Lemma 1 of Fan and Li (2011) with the additional parameter $\boldsymbol\eta$. 
However, this lemma can be proved in exactly the same way as in Fan and Li (2011) holding $\boldsymbol\eta$ constant throughout the proof and is omitted.

\begin{lemma}
Under the assumptions of Theorem \ref{THM:fixed_oracle}, for any given $\boldsymbol\beta^{(1)}$ satisfying $||\boldsymbol\beta^{(1)} - \boldsymbol\beta_0^{(1)}|| = O_P(n^{-1/2})$
and any $\boldsymbol\eta$ and constant $C$, we have with probability tending to one,
$$
Q_{n,\lambda}(\boldsymbol\beta^{(1)}, \boldsymbol{0}; \boldsymbol\eta) = \max_{||\boldsymbol\beta^{(2)}||\leq C n^{-1/2}}Q_{n,\lambda}(\boldsymbol\beta^{(1)}, \boldsymbol\beta^{(2)};\boldsymbol\eta).
$$
\label{LEM:fixed_oracle}
\end{lemma}

Now the first part of the theorem follows directly from the above lemma. 
The asymptotic normality of $\hat{\boldsymbol{\beta}}^{(1)}$ follows similarly as in the proof of Theorem 2 of Fan and Li (2011) using additionally the consistency of $\hat{\boldsymbol{\eta}}$
and the asymptotic distribution of $\hat{\boldsymbol{\eta}}$ which follows from the corresponding estimating equation,  in a standard way 
just as in the case of the usual MLE since there is no penalty involved in the estimation of $\boldsymbol\eta$.

\subsection{Proof of Theorem \ref{THM:consistency_high}}

The proof is an extension of the proof of Theorem B.1 of Fan and Liao (2014).\\
Let us denote $k_n=a_n + \sqrt{s}P_n'(d_n)$, which is $o(1)$ by our assumptions. 
Denote $Q_1(\boldsymbol\beta_S, \boldsymbol\eta) = Q_{n,\lambda}((\boldsymbol\beta_S^T, \boldsymbol{0})^T,\boldsymbol\eta)$.

Now, given any $\tau>0$, define the set 
$$
\mathcal{N}_\tau = \{(\boldsymbol\beta^T,\boldsymbol\eta^T)^T : \boldsymbol\beta\in \mathbb{R}^s, \boldsymbol\eta\in\mathbb{R}^d, 
||\boldsymbol\beta-\boldsymbol\beta_{S0}||\leq k_n\tau, ||\boldsymbol\eta-\boldsymbol\eta_{0}||\leq c_n\tau \},
$$
and denote its boundary by $\partial \mathcal{N}_\tau$ on which the inequalities becomes the equality.
To prove the first part of the theorem, we will fix an $\epsilon>0$ and show the existence of a $\tau>0$ such that 
$P(H_n(\tau)) > 1-\epsilon$ for all sufficiently large $n$, where the event $H_n(\tau)$ is defined as 
$$
H_n(\tau) = \{Q_1(\boldsymbol\beta_{S0}, \boldsymbol\eta_0) < \min_{(\boldsymbol\beta_S^T, \eta^T)^T \in \partial \mathcal{N}_\tau} Q_1(\boldsymbol\beta_S, \boldsymbol\eta) \}.
$$
This will suffice because, on the event $H_n(\tau)$, by the continuity of $Q_1(\boldsymbol\beta_S, \boldsymbol\eta)$, 
it follows that there exists a local minimizer $(\hat{\boldsymbol{\beta}}_S, \hat{\boldsymbol{\eta}})$ of $Q_1(\boldsymbol\beta_S, \boldsymbol\eta)$ inside $\mathcal{N}_\tau$ 
which gives the local minimizer $(\hat{\boldsymbol{\beta}}_S, 0, \hat{\boldsymbol{\eta}})$ as in the theorem.

Take any $(\boldsymbol\beta_S, \boldsymbol\eta)\in \partial\mathcal{N}_\tau$ which then satisfies $||\boldsymbol\beta_S-\boldsymbol\beta_{S0}||\leq k_n\tau$ and $||\boldsymbol\eta-\boldsymbol\eta_{0}||\leq c_n$.
Denote $\boldsymbol\phi=(\boldsymbol\beta_S,\boldsymbol\eta)$ and $\boldsymbol\phi_0=(\boldsymbol\beta_{S0}, \boldsymbol\eta_0)$.
By suitable Taylor series expansion we get a $(\boldsymbol\beta^*, \boldsymbol\eta^*)$ lying on the segment joining $(\boldsymbol\beta_S,\boldsymbol\eta)$ and $(\boldsymbol\beta_{S0},\boldsymbol\eta_0)$ such that
\begin{eqnarray}
Q_1(\boldsymbol\beta_S,\boldsymbol\eta) - Q_1(\boldsymbol\beta_{S0},\boldsymbol\eta_0) 
&=& (\boldsymbol\beta_S - \boldsymbol\beta_{S0})^T\nabla_S L_1(\boldsymbol\beta_S, \boldsymbol\eta) + (\boldsymbol\eta - \boldsymbol\eta_{0})^T\nabla_\eta L_1(\boldsymbol\beta_S, \boldsymbol\eta) 
\nonumber\\
&&+ (\boldsymbol\phi - \boldsymbol\phi_0)^T\nabla^2 L_1(\boldsymbol\beta^*, \boldsymbol\eta^*) (\boldsymbol\phi-\boldsymbol\phi_0) 
+ \sum_{j=1}^s\left[P_{n,\lambda}(|\beta_{S,j}|) - P_{n,\lambda}(|\beta_{S0,j}|)\right].~~
\end{eqnarray} 
Now, let us consider the following events
\begin{eqnarray}
H_{11} &=& \left\{ (\boldsymbol\beta_S - \boldsymbol\beta_{S0})^T\nabla_S L_1(\boldsymbol\beta_S, \boldsymbol\eta) \geq -C_1 ||\boldsymbol\beta_S - \boldsymbol\beta_{S0}||a_n\right\},
\nonumber\\
H_{12} &=& \left\{ (\boldsymbol\eta - \boldsymbol\eta_{S0})^T\nabla_\eta L_1(\boldsymbol\beta_S, \boldsymbol\eta) \geq - C_2 ||\boldsymbol\eta_S - \boldsymbol\eta_{S0}||c_n\right\},
\nonumber\\
H_{2} &=& \left\{ (\boldsymbol\phi - \boldsymbol\phi_0)^T\nabla^2 L_1(\boldsymbol\beta_{S0}, \boldsymbol\eta_0) (\boldsymbol\phi-\boldsymbol\phi_0) > C_\epsilon ||\boldsymbol\phi - \boldsymbol\phi_0||^2 \right\}, 
\nonumber\\
H_{3} &=& \left\{ ||\nabla^2 L_1(\boldsymbol\beta_{S0}, \boldsymbol\eta_0) - \nabla^2 L_1(\boldsymbol\beta^*, \boldsymbol\eta^*)||_F < \frac{C_\epsilon}{4} \right\}, 
\nonumber\\
H_{4} &=& \left\{ (\boldsymbol\phi - \boldsymbol\phi_0)^T\nabla^2 L_1(\boldsymbol\beta^*, \boldsymbol\eta^*) (\boldsymbol\phi-\boldsymbol\phi_0) 
> \frac{3C_\epsilon}{4} ||\boldsymbol\phi - \boldsymbol\phi_0||^2 \right\}.\nonumber
\end{eqnarray} 
Now, by Assumption L1(i), there exists $C_1$ and $C_2$ such that $P(H_{11})>1-\epsilon/8$ and  $P(H_{12})>1-\epsilon/8$ for all sufficiently large $n$
so that we have  $P(H_{1})>1-\epsilon/4$, where  we define $H_1 = H_{11}\cap H_{12}$.
Also, by Assumption L1(ii) and L1(iii), we get an $C_\epsilon$ satisfying $P(H_{2})>1-\epsilon/4$ 
and $P(H_{3})>1-\epsilon/4$ for all sufficiently large $n$ and for any $\tau>0$.
Finally, by noting that $H_4\subseteq H_2\cap H_3$, we have $P(H_{4})>1-\epsilon/2$ for all large $n$.

Further, by Lemma B.1 of Fan and Liao (2014), we get 
$$
\sum_{j=1}^s\left[P_{n,\lambda}(|\beta_{S,j}|) - P_{n,\lambda}(|\beta_{S0,j}|)\right] \geq - \sqrt(s) P_n'(d_n)||\boldsymbol\beta_S - \boldsymbol\beta_{S0}||
$$
and we also have $C_1 a_n + \sqrt{s}P_{n,\lambda}'(d_n) \leq (C_1+1)k_n$ by definition of $k_n$.
Combining all these, we have, on $H_1\cap H_4$,
\begin{eqnarray}
Q_1(\boldsymbol\beta_S,\boldsymbol\eta) - Q_1(\boldsymbol\beta_{S0},\boldsymbol\eta_0) &\geq& k_n\tau \left(\frac{3k_n\tau C_\epsilon}{8} - (C_1+1)k_n\right) + 
c_n\tau \left(\frac{3c_n\tau C_\epsilon}{8} - C_2 c_n\right)
\nonumber\\
&\geq & 0, 
\end{eqnarray} 
uniformly on $\partial\mathcal{N}_\tau$, by choosing $\tau > \frac{8\max\{C_1+1, C_2\}}{3C_\epsilon}$.
This completes the proof of the first part by noting that, under above choices,  
$P(H_n(\tau)) \geq P(H_1\cap H_4) \geq 1-\epsilon$ for all sufficiently large $n$.

Next, we need to show that the local minimizer in $\mathcal{N}_\tau$, denoted by $(\hat{\boldsymbol{\beta}}_S, \hat{\boldsymbol{\eta}})$ is strict with probability arbitrarily close to one.
Let us define, for $h\in \mathbb{R}-\{0\}$, 
\begin{equation}
\zeta_1(h) = \limsup_{\epsilon \rightarrow 0+} \sup_{t_1<t_2: (t_1,t_2)\in (|h|-\epsilon, |h|+\epsilon)}
- \left[\frac{P_{n,\lambda}(t_2)-P_{n,\lambda}(t_1)}{t_2-t_1}\right].
\label{EQ:zeta_h}
\end{equation}
Note that $\zeta_1(\cdot)\geq 0$ by concavity of the penalty $P_{n,\lambda}(\cdot)$ and $L_1$ is twice differentiable.
So, it is enough to show that $\boldsymbol{A}(\hat{\boldsymbol{\beta}}_S, \hat{\boldsymbol{\eta}})$ is positive definite, where 
$\boldsymbol{A}(\boldsymbol\beta_S,\boldsymbol\eta) = \nabla^2L_1(\boldsymbol\beta_S, \boldsymbol\eta) - {\rm Diag}\{\zeta_1(\beta_{S,1}), \cdots, \zeta_1(\beta_{S,s}), 0, \cdots, 0\}$.
Again, let us break down the problem through the following events 
\begin{eqnarray}
H_{5} &=& \left\{ \zeta(\hat{\boldsymbol{\beta}}_S) \leq \sup_{\boldsymbol\beta\in B(\boldsymbol\beta_{S0}, cd_n)}\zeta(\boldsymbol\beta) \right\},
\nonumber\\
H_{6} &=& \left\{ ||\nabla^2 L_1(\hat{\boldsymbol{\beta}}_{S}, \hat{\boldsymbol{\eta}}) - \nabla^2 L_1(\boldsymbol\beta_{S0}, \boldsymbol\eta_0)||_F < \frac{C_\epsilon}{4} \right\}, 
\nonumber\\
H_{7} &=& \left\{ \lambda_{\min} (\nabla^2 L_1(\boldsymbol\beta_{S0}, \boldsymbol\eta_0))> C_\epsilon \right\}.\nonumber
\end{eqnarray} 
Note that, on $H_5$, 
$\max_{j\leq s} \zeta_1(\hat{{\beta}}_{S,j}) \leq \zeta(\hat{\boldsymbol{\beta}}_S) \leq \sup_{\boldsymbol\beta\in B(\boldsymbol\beta_{S0}, cd_n)}\zeta(\boldsymbol\beta)$
and hence on $H_5 \cap H_6 \cap H_7$, we have for all sufficiently large $n$, for any $\boldsymbol\alpha\in \mathbb{R}^{s+q}$ satisfying $||\boldsymbol\alpha||=1$,
\begin{eqnarray}
\boldsymbol\alpha^T \boldsymbol{A}(\hat{\boldsymbol{\beta}}_S, \hat{\boldsymbol{\eta}}) \boldsymbol\alpha 
&\geq& \boldsymbol\alpha^T \nabla^2 L_1(\boldsymbol\beta_{S0}, \boldsymbol\eta_0)\boldsymbol\alpha 
- \left| \boldsymbol\alpha^T \left(\nabla^2 L_1(\hat{\boldsymbol{\beta}}_{S}, \hat{\boldsymbol{\eta}}) - \nabla^2 L_1(\boldsymbol\beta_{S0}, \boldsymbol\eta_0)\right)\alpha \right|
- \max_{j\leq s} \zeta_1(\hat{{\beta}}_{S,j}) 
\nonumber\\
&\geq& \frac{3C_\epsilon}{4} - \sup_{\boldsymbol\beta\in B(\boldsymbol\beta_{S0}, cd_n)}\zeta(\boldsymbol\beta)
\nonumber \\
&\geq& \frac{C_\epsilon}{4}, ~~~~ \mbox{ by Assumption P(iv).}\nonumber
\end{eqnarray}
This implies $\lambda_{\min}(\boldsymbol{A}(\hat{\boldsymbol{\beta}}_S, \hat{\boldsymbol{\eta}}) ) \geq \frac{C_\epsilon}{4}$ for all sufficiently large $n$.
However, we get from Assumption L1(ii) that $P(H_7)>1-\epsilon$. Finally, to show that $P(H_5 \cap H_6)>1-\epsilon$,
we note that
$$
P(H_5) \geq P(\hat{\boldsymbol{\beta}}_S\in B(\boldsymbol\beta_{S0}, cd_n)) \geq 1 - \epsilon/2, ~~~\mbox{ since } k_n = o(d_n),
$$
and 
\begin{eqnarray}
P(H_6^c) &\leq& P(H_6^c, ||\hat{\boldsymbol{\beta}}_{S} - \boldsymbol\beta_{S0}|| \leq k_n) + P(||\hat{\boldsymbol{\beta}}_{S} - \boldsymbol\beta_{S0}|| >  k_n) 
\nonumber\\
&\leq& P\left(\sup_{||\boldsymbol\beta_S - \boldsymbol\beta_{S0}||\leq\alpha_n, ||\boldsymbol\eta - \boldsymbol\eta_{0}||\leq\gamma_n}
||\nabla_{S\eta}^2 L(\boldsymbol\beta_{S}, \boldsymbol{0}; \boldsymbol\eta) - \nabla_{S\eta}^2 L(\boldsymbol\beta_{S0}, \boldsymbol{0}; \boldsymbol\eta_0)|| \geq C_\epsilon/4 \right) + \epsilon/4
\nonumber \\
&\leq& \frac{\epsilon}{2}.
\end{eqnarray}
This completes the proof.

\subsection{Proof of Theorem \ref{THM:variable_high}}
The proof follows by a direct extension of the proof of Theorem B.2 of Fan and Liao (2014) and 
using the consistency of the local minimizer $(\hat{\boldsymbol{\beta}}_S, \hat{\boldsymbol{\eta}})$ obtained in Theorem  \ref{THM:consistency_high}.\\

Consider $\hat{\boldsymbol{\beta}}=(\hat{\boldsymbol{\beta}}_S^T, \boldsymbol{0})^T$. We have to show that there is a random neighborhood $\mathcal{H}$, say, of $(\hat{\boldsymbol{\beta}}, \hat{\boldsymbol{\eta}})$
so that we have $Q_{n,\lambda}(\hat{\boldsymbol{\beta}}, \hat{\boldsymbol{\eta}}) < Q_{n,\lambda}(\boldsymbol\beta, \boldsymbol\eta)$ with probability tending to one 
for all $(\boldsymbol\beta,\boldsymbol\eta)\in\mathcal{H}$ with $\boldsymbol\beta=(\boldsymbol\beta_S,\boldsymbol\beta_N)^T$ and $\boldsymbol\beta_N\neq \boldsymbol{0}$. 
However, by definition of $\hat{\boldsymbol{\beta}}_S$, we can take $\mathcal{H}$ sufficiently small so that $Q_1(\hat{\boldsymbol{\beta}}_S,\hat{\boldsymbol{\eta}}) \leq Q_1(\boldsymbol\beta_S, \boldsymbol\eta)$
and hence we have $Q_{n,\lambda}(T\hat{\boldsymbol{\beta}},\hat{\boldsymbol{\eta}}) = Q_1(\hat{\boldsymbol{\beta}}_S,\hat{\boldsymbol{\eta}}) \leq Q_1(\boldsymbol\beta_S, \boldsymbol\eta) 
= Q_{n,\lambda}(T\boldsymbol\beta, \boldsymbol\eta)$.
Hence, it is enough to show that there is a sufficiently small neighborhood $\mathcal{H}$ of $(\hat{\boldsymbol{\beta}}, \hat{\boldsymbol{\eta}})$
so that we have $Q_{n,\lambda}(T\boldsymbol\beta, \boldsymbol\eta) < Q_{n,\lambda}(\boldsymbol\beta, \boldsymbol\eta)$ with probability tending to one 
for all $(\boldsymbol\beta,\boldsymbol\eta)\in\mathcal{H}$ with $\boldsymbol\beta=(\boldsymbol\beta_S,\boldsymbol\beta_N)^T$ and $\boldsymbol\beta_N\neq \boldsymbol{0}$. 
But, this follows directly from our Assumption (L2), since
\begin{eqnarray}
Q_{n,\lambda}(T\boldsymbol\beta, \boldsymbol\eta) - Q_{n,\lambda}(\boldsymbol\beta, \boldsymbol\eta) &=& 
L_n(T\boldsymbol\beta, \boldsymbol\eta) - L_n(\boldsymbol\beta, \boldsymbol\eta) - \left(\sum_{j=1}^p P_{n,\lambda}(\beta_j) - \sum_{j=1}^s P_{n,\lambda}((T\beta)_j)\right)
\nonumber \\
&<& 0.
\end{eqnarray}
This proves the first part (i) of the theorem. 

The second part (ii) of the theorem follows from the above inequality along with the second part of Theorem  \ref{THM:consistency_high}.

\begin{table*}[h]
	\centering
	\caption{Empirical mean, SD and MSE of the parameter estimates based on $L_1$ and SCAD penalty for low-dimensional set-up with different $\rho$,
		along with estimated active set size ($|S(\hat{\boldsymbol\beta})|$), number of true positives (TP) and 
		the model prediction error (PE) adjusted for the random effects 
		(the column $\boldsymbol\beta_N$ denotes the values corresponding to averaged $\beta_j$s for $j$ not in the true active set, 
		i.e., over $\beta_6$ to $\beta_p$)}
	\resizebox{.75\textwidth}{!}{
		\begin{tabular}{ll|cc|c|cccccc|cc|}\hline
			&		&	$|S(\hat{\boldsymbol\beta})|$	&	TP	&	PE	&	$\beta_1$	&	$\beta_2$	&	$\beta_3$	&	$\beta_4$	&	$\beta_5$	&	$\boldsymbol\beta_N$	&	$\sigma^2$	&	$\theta^2$	\\	\hline
			\multicolumn{13}{l|}{$L_1$ Penalty}																									\\	\hline
			\multicolumn{13}{c|}{$p=10$}																									\\	\hline
			$\rho=0$	&	Mean	&	6.01	&	5.00	&	0.17	&	0.99	&	2.02	&	3.97	&	2.97	&	2.97	&	0.00	&	0.23	&	0.41	\\	
			&	SD	&	1.19	&	0.00	&	0.02	&	0.32	&	0.37	&	0.05	&	0.06	&	0.06	&	0.03	&	0.03	&	0.22	\\	
			&	MSE	&		&		&		&	0.1044	&	0.1343	&	0.0038	&	0.0045	&	0.0039	&	0.0008	&	0.0012	&	0.0717	\\	\hline
			$\rho=0.5$	&	Mean	&	5.67	&	5.00	&	0.17	&	0.96	&	1.99	&	3.99	&	3.00	&	2.98	&	0.00	&	0.24	&	0.42	\\	
			&	SD	&	0.94	&	0.00	&	0.03	&	0.39	&	0.36	&	0.06	&	0.06	&	0.06	&	0.03	&	0.03	&	0.22	\\	
			&	MSE	&		&		&		&	0.1486	&	0.1252	&	0.0039	&	0.0037	&	0.0045	&	0.0007	&	0.0013	&	0.0663	\\	\hline
			\multicolumn{13}{c|}{$p=50$}																									\\	\hline
			$\rho=0$	&	Mean	&	8.07	&	5.00	&	0.17	&	0.96	&	1.99	&	3.97	&	2.95	&	2.95	&	0.00	&	0.23	&	0.46	\\	
			&	SD	&	2.62	&	0.00	&	0.03	&	0.33	&	0.41	&	0.05	&	0.06	&	0.06	&	0.02	&	0.04	&	0.26	\\	
			&	MSE	&		&		&		&	0.1095	&	0.1693	&	0.0040	&	0.0063	&	0.0061	&	0.0003	&	0.0022	&	0.0761	\\	\hline
			$\rho=0.5$	&	Mean	&	7.99	&	5.00	&	0.17	&	0.92	&	2.03	&	3.97	&	3.01	&	2.98	&	0.00	&	0.23	&	0.37	\\	
			&	SD	&	2.55	&	0.00	&	0.03	&	0.34	&	0.35	&	0.06	&	0.06	&	0.06	&	0.02	&	0.04	&	0.22	\\	
			&	MSE	&		&		&		&	0.1233	&	0.1216	&	0.0052	&	0.0034	&	0.0043	&	0.0003	&	0.0018	&	0.0831	\\	\hline
			\multicolumn{13}{l|}{SCAD Penalty}																									\\	\hline
			\multicolumn{13}{c|}{$p=10$}																									\\	\hline
			$\rho=0$	&	Mean	&	5.23	&	5.00	&	0.17	&	1.05	&	2.02	&	4.00	&	3.01	&	3.00	&	0.00	&	0.24	&	0.40	\\	
			&	SD	&	0.53	&	0.00	&	0.03	&	0.35	&	0.34	&	0.05	&	0.05	&	0.05	&	0.02	&	0.03	&	0.25	\\	
			&	MSE	&		&		&		&	0.1203	&	0.1181	&	0.0025	&	0.0029	&	0.0025	&	0.0003	&	0.0012	&	0.0849	\\	\hline
			$\rho=0.5$	&	Mean	&	5.22	&	5.00	&	0.18	&	0.99	&	1.98	&	3.99	&	2.99	&	3.01	&	0.00	&	0.25	&	0.43	\\	
			&	SD	&	0.64	&	0.00	&	0.03	&	0.30	&	0.34	&	0.06	&	0.06	&	0.06	&	0.02	&	0.04	&	0.21	\\	
			&	MSE	&		&		&		&	0.0892	&	0.1136	&	0.0034	&	0.0042	&	0.0033	&	0.0004	&	0.0013	&	0.0596	\\	\hline
			\multicolumn{13}{c|}{$p=50$}																									\\	\hline
			$\rho=0$	&	Mean	&	5.69	&	5.00	&	0.17	&	1.02	&	2.03	&	4.00	&	3.01	&	3.00	&	0.00	&	0.24	&	0.43	\\	
			&	SD	&	1.35	&	0.00	&	0.03	&	0.36	&	0.39	&	0.05	&	0.04	&	0.05	&	0.01	&	0.04	&	0.21	\\	
			&	MSE	&		&		&		&	0.1275	&	0.1512	&	0.0025	&	0.0018	&	0.0025	&	0.0001	&	0.0016	&	0.0602	\\	\hline
			$\rho=0.5$	&	Mean	&	5.59	&	5.00	&	0.17	&	1.01	&	2.04	&	4.00	&	2.99	&	3.00	&	0.00	&	0.24	&	0.41	\\	
			&	SD	&	1.16	&	0.00	&	0.03	&	0.35	&	0.38	&	0.06	&	0.07	&	0.06	&	0.01	&	0.04	&	0.22	\\	
			&	MSE	&		&		&		&	0.1243	&	0.1440	&	0.0042	&	0.0049	&	0.0040	&	0.0001	&	0.0014	&	0.0699	\\	\hline
		\end{tabular}}
		\label{TAB:Sim_L1}
	\end{table*}

	\begin{table*}[h]
		\centering
		\caption{Empirical mean, SD and MSE of the parameter estimates based on $L_1$ and SCAD penalty for high-dimensional set-up with different $\rho$,
			along with estimated active set size ($|S(\hat{\boldsymbol\beta})|$), number of true positives (TP) and 
			the model prediction error (PE) adjusted for the random effects 
			(the column $\boldsymbol\beta_N$ denotes the values corresponding to averaged $\beta_j$s for $j$ not in the true active set, 
			i.e., over $\beta_6$ to $\beta_p$)}
		\resizebox{.8\textwidth}{!}{
			\begin{tabular}{ll|cc|c|cccccc|cc|}\hline
				&		&	$|S(\hat{\boldsymbol\beta})|$	&	TP	&	PE	&	$\beta_1$	&	$\beta_2$	&	$\beta_3$	&	$\beta_4$	&	$\beta_5$	&	$\boldsymbol\beta_N$	&	$\sigma^2$	&	$\theta^2$	\\	\hline
				\multicolumn{13}{l|}{$L_1$ Penalty}																									\\	\hline
				\multicolumn{13}{c|}{$p=300$}																									\\	\hline
				$\rho=0$	&	Mean	&	11.23	&	5.00	&	0.15	&	1.02	&	2.02	&	3.94	&	2.95	&	2.96	&	0.00	&	0.21	&	0.39	\\	
				&	SD	&	4.24	&	0.00	&	0.03	&	0.34	&	0.32	&	0.05	&	0.06	&	0.06	&	0.01	&	0.04	&	0.22	\\	
				&	MSE	&		&		&		&	0.1149	&	0.1019	&	0.0057	&	0.0061	&	0.0050	&	0.0001	&	0.0029	&	0.0772	\\	\hline
				$\rho=0.5$	&	Mean	&	9.37	&	5.00	&	0.16	&	1.01	&	1.97	&	3.97	&	2.98	&	2.96	&	0.00	&	0.22	&	0.44	\\	
				&	SD	&	3.79	&	0.00	&	0.03	&	0.35	&	0.40	&	0.07	&	0.07	&	0.07	&	0.01	&	0.04	&	0.25	\\	
				&	MSE	&		&		&		&	0.1227	&	0.1594	&	0.0059	&	0.0049	&	0.0059	&	0.0001	&	0.0024	&	0.0757	\\	\hline
				\multicolumn{13}{c|}{$p=500$}	\\	\hline
				$\rho=0$	&	Mean	&	10.85	&	5.00	&	0.15	&	0.97	&	1.95	&	3.93	&	2.94	&	2.94	&	0.00	&	0.22	&	0.42	\\	
				&	SD	&	4.12	&	0.00	&	0.03	&	0.36	&	0.33	&	0.05	&	0.06	&	0.06	&	0.01	&	0.04	&	0.24	\\	
				&	MSE	&		&		&		&	0.1278	&	0.1096	&	0.0078	&	0.0066	&	0.0075	&	0.0001	&	0.0032	&	0.0751	\\	\hline
				$\rho=0.5$	&	Mean	&	10.53	&	5.00	&	0.16	&	1.05	&	2.05	&	3.97	&	2.98	&	2.96	&	0.00	&	0.22	&	0.36	\\	
				&	SD	&	3.89	&	0.00	&	0.03	&	0.40	&	0.38	&	0.08	&	0.07	&	0.07	&	0.01	&	0.04	&	0.23	\\	
				&	MSE	&		&		&		&	0.1625	&	0.1475	&	0.0065	&	0.0051	&	0.0058	&	0.0001	&	0.0026	&	0.0893	\\	\hline
				\multicolumn{13}{l|}{SCAD Penalty}																									\\	\hline
				\multicolumn{13}{c|}{$p=300$}																									\\	\hline
				$\rho=0$	&	Mean	&	7.29	&	5.00	&	0.16	&	1.06	&	2.00	&	3.99	&	3.00	&	3.00	&	0.00	&	0.22	&	0.42	\\	
				&	SD	&	3.57	&	0.00	&	0.03	&	0.37	&	0.36	&	0.05	&	0.05	&	0.05	&	0.01	&	0.04	&	0.24	\\	
				&	MSE	&		&		&		&	0.1403	&	0.1279	&	0.0029	&	0.0027	&	0.0026	&	0.0001	&	0.0022	&	0.0745	\\	\hline
				$\rho=0.5$	&	Mean	&	7.22	&	5.00	&	0.16	&	0.98	&	2.00	&	4.01	&	3.00	&	2.99	&	0.00	&	0.22	&	0.43	\\	
				&	SD	&	3.58	&	0.00	&	0.03	&	0.37	&	0.33	&	0.06	&	0.06	&	0.06	&	0.01	&	0.04	&	0.24	\\	
				&	MSE	&		&		&		&	0.1368	&	0.1078	&	0.0042	&	0.0037	&	0.0038	&	0.0001	&	0.0021	&	0.0735	\\	\hline
				\multicolumn{13}{c|}{$p=500$}		\\	\hline
				$\rho=0$	&	Mean	&	8.30	&	5.00	&	0.15	&	1.04	&	1.94	&	3.99	&	2.99	&	3.00	&	0.00	&	0.21	&	0.42	\\	
				&	SD	&	4.16	&	0.00	&	0.03	&	0.33	&	0.34	&	0.05	&	0.05	&	0.06	&	0.00	&	0.04	&	0.29	\\	
				&	MSE	&		&		&		&	0.1093	&	0.1189	&	0.0024	&	0.0025	&	0.0033	&	0.0001	&	0.0031	&	0.1032	\\	\hline
				$\rho=0.5$	&	Mean	&	7.62	&	5.00	&	0.16	&	1.01	&	2.01	&	4.00	&	3.00	&	2.99	&	0.00	&	0.23	&	0.42	\\	
				&	SD	&	3.52	&	0.00	&	0.03	&	0.34	&	0.37	&	0.08	&	0.07	&	0.07	&	0.00	&	0.04	&	0.22	\\	
				&	MSE	&		&		&		&	0.1130	&	0.1350	&	0.0056	&	0.0051	&	0.0044	&	0.0000	&	0.0025	&	0.0696	\\	\hline
			\end{tabular}}
			\label{TAB:Sim_SCAD}
		\end{table*}

\begin{table*}
	\centering
	\caption{Estimated fixed effect coefficients ($\hat{\beta}$) for the real data set under mixed model and regression set-up 
		(their rank is given in the parenthesis).}
	\resizebox{0.6\textwidth}{!}{
		\begin{tabular}{l|cc|cc|}\hline
			&\multicolumn{2}{c|}{Mixed Model}  & \multicolumn{2}{c|}{Regression Model} \\
			Penalty 	&	SCAD		&	$L_1$		&	SCAD		&	$L_1$		\\\hline
			\multicolumn{5}{c}{Number of Genes Selected}		\\\hline
			&	29		&	30		&	32		&	37		\\\hline
			\multicolumn{5}{c}{Coefficients of Selected Genes}		\\\hline
			DOCK10 (*)	&	3.03	(1)	&	5.71	(1)	&	3.04	(1)	&	4.94	(1)	\\
			CAST (*)	&	2.73	(2)	&	3.02	(2)	&	2.83	(2)	&	3.18	(2)	\\
			GZMK (*)	&	2.43	(3)	&	0.26	(14)	&	1.84	(3)	&	1.43	(5)		\\
			NA 	&	2.08	(4)	&	1.68	(3)	&	1.82	(4)	&	2.41	(3)		\\
			HLA-H (*)	&	1.56	(5)	&	0.88	(8)	&	1.47	(6)	&	1.39	(6)		\\
			SLC22A16	&	1.52	(6)	&	--	(15)	&	0.58	(11)	&	0.85	(10)		\\
			GSTM1 (*)	&	1.38	(7)	&	0.91	(7)	&	1.55	(5)	&	1.35	(7)		\\
			NA 	&	1.13	(8)	&	0.31	(13)	&	0.86	(7)	&	0.71	(11)		\\
			SNX29	&	0.96	(9)	&	1.41	(4)	&	0.63	(9)	&	1.90	(4)		\\
			UTS2 (*)	&	0.73	(10)	&	0.48	(12)	&	0.59	(10)	&	0.45	(13)	\\
			FAM45A	&	0.34	(11)	&	1.09	(6)	&	0.15	(13)	&	0.96	(9)		\\
			LOC554223	&	0.26	(12)	&	0.59	(10)	&	0.56	(12)	&	0.68	(12)		\\
			ACCS	&	--	(13)	&	1.38	(5)	&	0.78	(8)	&	1.16	(8)	\\
			PJA2	&	--	(13)	&	0.63	(9)	&	--	(15)	&	0.30	(15)	\\
			NFIB	&	--	(13)	&	0.49	(11)	&	--	(15)	&	0.42	(14)		\\
			IRF5 (*)	&	--	(13)	&	--	(15)	&	0.05	(14)	&	0.10	(16)	\\
			LOC100170939	&	--	(13)	&	--	(15)	&	--	(15)	&	-0.02	(17)		\\
			MYL4	&	--	(13)	&	--	(15)	&	-0.17	(21)	&	-0.06	(19)		\\
			PKIA	&	--	(13)	&	-0.57	(25)	&	--	(15)	&	-0.86	(25)		\\
			FGD2	&	--	(13)	&	-0.69	(26)	&	--	(15)	&	-0.05	(18)		\\
			MX1 (*)	&	-0.12	(21)	&	-0.52	(24)	&	-0.40	(22)	&	-0.47	(21)		\\
			HSH2D (*)	&	-0.80	(22)	&	-0.40	(23)	&	-0.52	(23)	&	-0.86	(24)		\\
			LOC644936	&	-1.02	(23)	&	-1.34	(30)	&	-1.18	(27)	&	-1.07	(26)		\\
			PPAT	&	-1.21	(24)	&	-0.97	(28)	&	-1.03	(24)	&	-0.81	(23)			\\
			NA 	&	-1.23	(25)	&	-0.77	(27)	&	-1.08	(25)	&	-0.80	(22)	\\
			NAPRT1	&	-1.36	(26)	&	-1.60	(31)	&	-1.60	(30)	&	-1.59	(29)		\\
			N4BP2L2	&	-1.49	(27)	&	-1.92	(34)	&	-1.82	(32)	&	-1.75	(32)		\\
			GYPC (*)	&	-1.63	(28)	&	-0.07	(22)	&	-1.61	(31)	&	-1.29	(27)		\\
			CENPK	&	-1.66	(29)	&	-1.66	(32)	&	-1.51	(29)	&	-1.69	(30)	\\
			COL18A1	&	-1.95	(30)	&	-1.16	(29)	&	-1.39	(28)	&	-1.44	(28)	\\
			C1orf85 (*)	&	-1.98	(31)	&	--	(15)	&	-0.10	(20)	&	-0.15	(20)	\\
			ZNF266	&	-2.09	(32)	&	--	(15)	&	-2.51	(35)	&	-2.60	(35)	\\
			COMMD2 (*)	&	-2.26	(33)	&	-2.42	(35)	&	-2.28	(34)	&	-2.28	(34)	\\
			ANPEP	&	-2.27	(34)	&	-1.70	(33)	&	-1.94	(33)	&	-2.01	(33)	\\
			PRUNE2	&	-2.91	(35)	&	--	(15)	&	-1.14	(26)	&	-1.72	(31)	\\
			NAIP (*)	&	-2.96	(36)	&	-2.77	(36)	&	-3.20	(36)	&	-3.58	(36)	\\
			PKIA	&	-4.07	(37)	&	-4.72	(37)	&	-4.19	(37)	&	-4.68	(37)	\\
			\hline
		\end{tabular}}
		\label{TAB:real_beta}
	\end{table*}

\begin{table*}
	\centering
	\caption{Estimated variance component parameters for the real data set under mixed model set-up.}
	\resizebox{!}{!}{
		\begin{tabular}{l|ccc|}\hline
			Penalty	& $\hat{\sigma}$ &$\hat{\theta}_I$ &$\hat{\theta}_{Time}$ \\\hline
			SCAD	& 3.134	& 0	& 0.520 \\
			$L_1$ &	3.435 &	0	& 0.571 \\\hline
		\end{tabular}}
		\label{TAB:real_eta}
	\end{table*}

\end{document}